\documentclass[12pt]{article}
\usepackage{makeidx,color}

\usepackage{epsfig}
\usepackage{inputenc}
\usepackage{my}
\inputencoding{latin1}
\newcommand{\ga}{\gamma}

\def\prp{\perp}
\def\Prp{T}
\def\sx{small-$x$}
\def\kt{\ensuremath{k_\prp}}
\def\kti#1{\ensuremath{k_{\prp #1}}}
\def\pt{\ensuremath{p_\prp}}
\def\pti#1{\ensuremath{p_{\prp #1}}}

\def\qti#1{\ensuremath{q_{\prp #1}}}
\def\sub#1{\ensuremath{_{\mbox{\scriptsize #1}}}}

\def\alb{\ensuremath{\bar{\alpha}\sub{S}}}

\def\ariadne{{\sc Ariadne}}

\def\jetset{{\sc Jetset}}
\def\pythia{{\sc Pythia}}

\def\ldcmc{{\small LDCMC}}

\def\as{\ensuremath{\alpha_s}}
\def\ord#1{\ensuremath{{\cal O}(#1)}}
\def\ordas{\ord{\as}}

\def\laeq{\,\lower3pt\hbox{$\buildrel < \over\sim$}\,}

\newcounter{Aenumct}

\renewcommand{\descriptionlabel}[1]%
{\code{#1}\hspace{-2mm}}

\newcommand{\CCFM}{CCFMa,CCFMb,CCFMc,CCFMd}
\newcommand{\BFKL}{BFKLa,BFKLb,BFKLc}
\newcommand{\LDCMC}{LDCa,LDCb,LDCc,LDCd}
\newcommand{\alphasb}{\alb}

\newcommand{\PYTHIAMC}{Jetsetc}

\newcommand{\DGLAP}{DGLAPa,DGLAPb,DGLAPc,DGLAPd}
\newcommand{\asb}{\alb}
\newcommand{\SMALLXMC}{SMALLXa,SMALLXb}
\newcommand{\CASCADEMC}{jung_salam_2000,CASCADE,CASCADEMC}

\newcommand{\BFKLNLO}{NLLFL,ciafaloni-camici}
\newcommand{\PT}{Dokshitzer:1978yd,Parisi:1979se,Collins:1985kg,Collins:1977iv}
\def\CASCADE{{\sc Cascade}}
\def\SMALLX{{\sc Smallx}}

\newcommand{\Pmax}{\bar{q}}
\newcommand{\cA}{{\cal A}}
\newcommand{\offshellmeqq}{CCH,CE}

\begin{document}

\begin{titlepage}

  \renewcommand{\thefootnote}{\arabic{footnote}}

  \begin{flushright}
   DESY 02-041\\
    hep-ph/0204115\\
  \end{flushright}
  \begin{center}
    
    \vskip 10mm \textbf{
     {\LARGE {\boldmath Small $x$ } Phenomenology \\
    Summary and Status  }}
    \vskip 15mm {\large
    
    {\sc The Small x Collaboration } \\ \vskip 5mm 
      Bo~Andersson\footnote[1]{Department of Theoretical Physics, Lund
        University, Sweden},
      Sergei~Baranov\footnote[2]{Lebedev~Institute~of~Physics, 
         Moscow, Russia},
      Jochen~Bartels\footnote[3]{II.\ Institut f\"ur Theoretische Physik,
        Universit\"at Hamburg, Germany},
      Marcello~Ciafaloni\footnote[4]{Dipartimento di Fisica, 
	Universit\`a di Firenze and INFN - Sezione di Firenze,
      Firenze (Italy)},
      John~Collins\footnote[5]{Penn State Univ., 104 Davey Lab.,
    University Park PA 16802, USA.},
      Mattias~Davidsson\footnote[6]{Department of Physics, Lund University,
        Sweden},
      Gösta~Gustafson\footnotemark[1],
      Hannes~Jung\footnotemark[6],
      Leif~Jönsson\footnotemark[6],
      Martin~Karlsson\footnotemark[6],
      Martin~Kimber\footnote[7]{Department of Physics, University of Durham, 
	Durham, UK},
      Anatoly Kotikov\footnote[8]{Bogoliubov Theoretical Physics Laboratory,
      Joint Institute for Nuclear Research, Dubna, Russia.},
      Jan~Kwiecinski\footnote[9]{
	H. Niewodniczanski Institute of Nuclear Physics, Krakow, Poland},
      Leif~Lönnblad\footnotemark[1],
      Gabriela~Miu\footnotemark[1],
      Gavin~Salam\footnote[10]{LPTHE, Universit\'es P. \& M. Curie 
	(Paris VI) et Denis Diderot
     (Paris VII), Paris, France},
      Mike~H.~Seymour\footnote[11]{Theoretical Physics, Department
      of Physics and Astronomy, University of Manchester, UK},
      Torbjörn~Sjöstrand\footnotemark[1],
      Nikolai~Zotov\footnote[12]{Skobeltsyn~Institute~of~Nuclear~Physics,  
      Moscow~State~University, Moscow, Russia}}

  \end{center}
  \vskip 15mm
  \begin{abstract}
    
    The aim of this paper is to summarize the general status of our
    understanding of \sx\  physics. It is based on presentations
    and discussions at an informal meeting on this topic held in Lund,
    Sweden, in March 2001.
    
    This document also marks the founding of an informal collaboration
    between experimentalists and theoreticians with a special interest
    in \sx\ physics.

    This paper is dedicated to the memory of Bo Andersson, who died 
    unexpectedly from a heart attack on March 4th, 2002.
    
  \end{abstract}

\end{titlepage}

\section{Introduction}
\label{sec:intro}
In this paper we present
a summary of the workshop on \sx\ parton dynamics held in Lund in
the beginning of March 2001. During two days we went through a number
of theoretical and phenomenological aspects of \sx\ physics in short
talks and long discussions. Here we will present the main points
of these discussions and try to summarize the general status of the
work in this field.

For almost thirty years, QCD has been the theory of strong
interactions. Although it has been very successful, there are still a
number of problems which have not been solved. Most of these have to
do with the transition between the perturbative and non-perturbative
description of the theory. Although perturbative techniques work
surprisingly well down to very small scales where the running coupling
starts to become large, in the end what is observed are hadrons, the
transition to which is still not on firm theoretical grounds. At very
high energies another problem arises. Even at high scales where the
running coupling is small the phase space for additional emissions
increases rapidly and makes the perturbative expansion ill-behaved.
The solution to this problem is to resum the leading logarithmic
behavior of the cross section to all orders, thus rearranging the
perturbative expansion into a more rapidly converging series.

The DGLAP\cite{\DGLAP} evolution is the most familiar
resummation strategy. Given that a cross section involving incoming
hadrons is dominated by diagrams where successive emissions are
strongly ordered in virtuality, the resulting large logarithms of
ratios of subsequent virtualities can be resummed. The cross section
can then be rewritten in terms of a process-dependent hard matrix
element convoluted with universal parton density functions, the
scaling violations of which are described by the DGLAP evolution. This is
called collinear factorization. Because of the strong ordering of
virtualities, the virtuality of the parton entering the hard
scattering matrix element can be neglected (treated collinear with the
incoming hadron) compared to the large scale $Q^2$.  This approach has been very
successful in describing the bulk of experimental measurements at
lepton--hadron and hadron--hadron colliders.

With HERA, a new kinematic regime has opened up where the very small
$x$ parts of the proton parton distributions are being probed. The hard
scale, $Q^2$, is not very high in such events and it was expected that the
DGLAP evolution should break down. To some surprise, the DGLAP evolution
has been quite successful in describing the strong rise of the cross
section with decreasing $x$. For some non-inclusive observables there
are, however, clear discrepancies as summarized below in
table~\ref{tab:collfac}.
\par
At asymptotically large energies, it is believed that the
theoretically correct description is given by the BFKL~\cite{\BFKL}
evolution. Here, each emitted gluon is assumed to take a large
fraction of the energy of the propagating
gluon, $(1-z)$ for $z \to 0$, 
and large logarithms of $1/z$ are summed up to all orders.
\begin{table}[htb]
  \begin{center}
    \begin{tabular}{|l|l|l|}
      \hline
      & collinear & \kt-\\
      & factorization & factorization\\
      \hline
      \hline
      HERA observables & & \\
      \hline
      \hline
      high $Q^2$ D$^*$ production & OK~\protect\cite{ZEUS_F2charm,H1_f2charm_2000} 
        & OK~\protect\cite{H1_f2charm_2000,Baranov2002}\\
      low $Q^2$ D$^*$ production & OK~\protect\cite{ZEUS_F2charm,H1_f2charm_2000} 
        & OK~\protect\cite{H1_f2charm_2000,Baranov2002}\\
      \hline
      direct photoproduction of D$^*$ & 1/2~\protect\cite{ZEUS_dstar} & 
        OK~\protect\cite{Baranov2002,baranov_zotov_1999,Charm,jung_salam_2000,baranov_zotov_2000}\\
        
      resolved photoproduction of D$^*$ & NO~\protect\cite{ZEUS_dstar} & 
        1/2~\protect\cite{baranov_zotov_1999,Charm,jung_salam_2000,baranov_zotov_2000}\\
      \hline
      high $Q^2$ B production & NO~\protect\cite{H1_bbar_dis} & ?\\
      low $Q^2$ B production & NO~\protect\cite{H1_bbar_dis} & ?\\
      \hline
      direct photoproduction of B& OK?~\protect\cite{ZEUS_bbar}, 
	NO~\protect\cite{H1_bbar} 
        & OK~\protect\cite{LipSalZot2000,jung_ringberg2001,jung-hq-2001} \\
      resolved photoproduction of B & OK?~\protect\cite{ZEUS_bbar}
        & OK~\protect\cite{LipSalZot2000,jung_ringberg2001,jung-hq-2001}\\
      \hline
      high $Q^2$ di-jets & OK~\protect\cite{ZEUS_resgamma_dis,H1_2+1jets_data} & ?\\
      low $Q^2$ di-jets & NO~\protect\cite{ZEUS_resgamma_dis,H1_2+1jets_data,ZEUS_resgamma_99,H1_resgamma_00} & ?\\
      \hline
      direct photoproduction of di-jets &
        1/2~\protect\cite{ZEUS_resgamma_dis,ZEUS_resgamma_99,H1_resgamma_00} & ?\\
      resolved photoproduction of di-jets & 
        NO~\protect\cite{ZEUS_resgamma_dis,ZEUS_resgamma_99,H1_resgamma_00} &
        ?\\
      \hline
      \hline
      HERA small-$x$ observables & & \\
      \hline
      \hline
      forward jet production & NO~\protect\cite{H1_fjets_data} 
        & OK~\protect\cite{jung_salam_2000}\\
      forward $\pi$ production & NO~\protect\cite{H1_fjets_data} & 1/2~\protect\cite{Martin} \\
      particle spectra & NO~\protect\cite{H1_energyflow} & OK~\protect\cite{jung_salam_2000}\\
      energy flow & NO~\protect\cite{H1_energyflow} & ?\\
      \hline
      photoproduction of $J/\Psi$ & NO~\protect\cite{H1_jpsi} & 
	1/2~\protect\cite{LiZo2000,Jung-ascona}\\
      $J/\Psi$ production in DIS & NO & ?\\
      \hline
      \hline
      TEVATRON observables & &\\
      \hline
      \hline
      high-$p_\prp$  D$^*$ production & ? & ?\\
      low-$p_\prp$  D$^*$ production & ? & ?\\
      \hline
      high-$p_\prp$  B production & OK?~\protect\cite{rfield}
	 & OK~\protect\cite{jung_ringberg2001,jung-hq-2001,Hagler_bbar,LiSaZo}\\
      low-$p_\prp$  B production & OK?~\protect\cite{rfield} 
	& OK~\protect\cite{jung_ringberg2001,jung-hq-2001,Hagler_bbar,LiSaZo}\\
      \hline
      $J/\Psi$ production & NO & ? \\
      high-$p_\prp$ jets at large rapidity differences & NO & ?\\
      \hline
    \end{tabular}
    
    \caption{{\it Summary of the ability of the collinear and
        \kt-factorization approaches to reproduce 
	  the current measurements of some observables: OK
        means a satisfactory description; 1/2 means a not perfect but 
          also a not too bad description, 
	    or in part of the phase space an acceptable
          description;
	    OK? means satisfactory description if a heavy quark
          excitation component is added in leading order; 
	    NO means that the 
        description is bad; and ? means that no thorough comparison has
        been made.}}
    \label{tab:collfac}
  \end{center}
\end{table}
Although the rise of $F_2$ with decreasing $x$ as measured at HERA
can be described with the DGLAP evolution, a strong power-like rise was 
predicted by BFKL.  Just as for DGLAP, it is possible to factorize an
observable into a convolution of process-dependent hard matrix
elements with universal parton distributions. But as the virtuality and
transverse momentum of the propagating gluon are no longer ordered, the
matrix elements have to be taken off-shell and the convolution is also
over transverse momentum with \emph{unintegrated} parton distributions.
We therefore talk about 
\kt-factorization~\cite{CCH,CE} or the semihard
approach~\cite{GLR,LRSS2}.
\par
Recently, the next-to-leading logarithmic (NLL) corrections to the
BFKL equation were calculated and found to be huge~\cite{\BFKLNLO}. 
This is related to
the fact that at any finite energy, the cross section will also get
contributions from emissions of gluons which take only a small
fraction of the energy of the propagating gluon.
\par
The CCFM~\cite{\CCFM}
 evolution equation resums also large logarithms of $1/(1-z)$
in addition to the $1/z$ ones. Furthermore it introduces angular
ordering of emissions to correctly treat gluon coherence effects. In
the limit of asymptotic energies, it is almost equivalent to 
BFKL~\cite{Forshaw:1998uq,Webber:1998we,Salam:1999ft}, but 
also similar to the DGLAP evolution for large $x$ and high
$Q^2$.  The cross section is still \kt-factorized into an off-shell
matrix element convoluted with an unintegrated parton density, which
now also contains a dependence on the maximum angle allowed in
emissions.
\par
An advantage of the CCFM evolution, compared to the BFKL evolution,
 is that it is fairly well
suited for implementation into an event generator program, which makes
quantitative comparison with data feasible also for non-inclusive
observables. There exist today three such 
generators~\cite{jung_salam_2000,SMALLXa,SMALLXb,%
  LDCa,LDCb,LDCc,LDCd,CASCADE,CASCADEMC} and they are
all being maintained or/and 
developed by people from the departments of
physics and of theoretical physics at Lund University.
\par
Since 1998 there has been an on-going project in Lund, supported by
the Swedish Royal Academy of Science, trying to get a better
understanding of the differences between the different generators and
to compare them to measured data.
This project is what led up to the meeting in Lund in early
March 2001, where a number of experts in the field were invited
to give short presentations and to discuss the current issues in \sx\ 
physics in general and \kt-factorization in particular.  
\par
In this article we will try to summarize these discussions and give a
general status report of this field of research. We also suggest the
formation of an informal collaboration of researchers in the field, to
facilitate a coherent effort to solve some of the current problems in
\sx\ parton dynamics.
\par
The outline of this article is as follows. First we give a general
introduction to \kt-factorization in
section~\ref{sec:kt-factorization}. Then, in section
\ref{sec:shell-matr-elem}, we discuss the off-shell matrix elements,
both at leading order and the prospects of going to next-to-leading
order. In section \ref{sec:un-integrated-parton}, we discuss the
unintegrated parton distributions and how they are evolved. Here we also
try to quantify the transition between DGLAP and BFKL. We present a
number of parameterizations of unintegrated parton distributions and make
a few comparisons. Then we describe the next-to-leading logarithmic
correction to the evolution. In section \ref{sec:gener-sc-evol} we
describe the available event generators for \sx\ evolution.
Finally in section \ref{sec:sx-collaboration} we
present our conclusions and discuss the forming of an informal
collaboration to better organize the future investigations of \sx\ 
phenomenology.

\section{The {\boldmath \kt}-factorization approach}
\label{sec:kt-factorization}

\begin{figure}[t]
  \begin{center}
    \begin{picture}(300,200)
      \put(30,0){\epsfig{figure=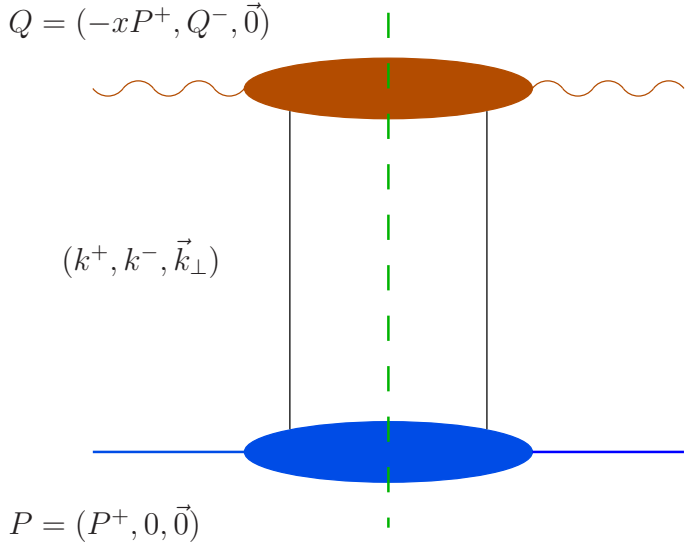,width=8cm}}
      \put(0,190){$Q=(-xP^+, Q^-, \vec{0})$}
      \put(0,0){$P=(P^+, 0, \vec{0})$}
      \put(20,100){$(k^+,k^-,\vec{k}_\prp)$}
    \end{picture}
    \caption{{\it Schematic picture of a typical unitarity diagram
        for deep inelastic scattering. An incoming proton with a large
        positive light-cone momentum, $P^+$, is being probed by a
        photon with a large virtuality and a large negative light-cone
        momentum. The photon scatters on a parton from the proton with
        space-like momentum $k$.}}
   \label{fig:unitarity}
  \end{center}
\end{figure}

\begin{figure}[htb]
\begin{center} 
\input{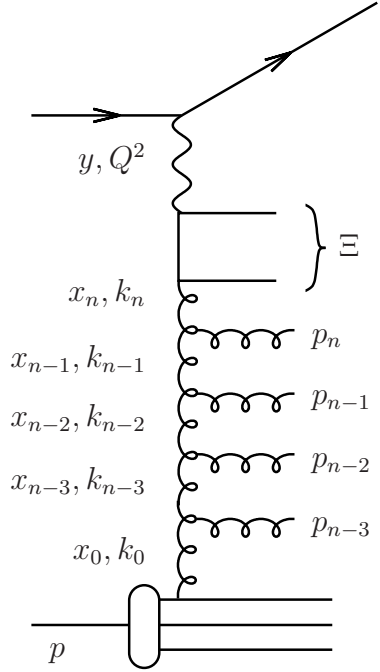}
\end{center} 
\caption{{\it Kinematic variables for multi-gluon emission. The
    $t$-channel gluon momenta are given by $k_i$ and the gluons
    emitted in the initial state cascade have momenta $p_i$. The upper
    angle for any emission is obtained from the quark box, as
    indicated with $\Xi$. We define $z_{\pm i}=k_{\pm i}/k_{\pm (i\mp
      1)}$ and $q_i=p_{\prp i}/(1-z_{+i})$.
    \label{fig:variables} }} 
\end{figure}
The calculation of inclusive quantities, like the structure function
$F_2(x,Q^2)$ at HERA, performed in NLO QCD is in perfect agreement with the
measurements.  
The NLO approach,
although phenomenologically successful for $F_2(x,Q^2)$, is
not fully satisfactory from a theoretical viewpoint because,
in the words of Catani, 
{\it ``the truncation of the
splitting functions at a fixed perturbative order is equivalent to assuming that
the dominant dynamical mechanism leading to scaling violations is the evolution
of parton cascades with strongly ordered transverse 
momenta"}~\cite{catani-feb2000}.
\par
As soon as exclusive quantities like jet or heavy quark
production  are investigated, the agreement between NLO coefficient functions
convoluted with NLO DGLAP
 parton distributions and the data is not at all
satisfactory. Large so-called $K$-factors 
(normalization factors, for example $K=\frac{\sigma_{tot}}{\sigma_{NLO}}$)
 are needed
to bring the NLO calculations close to the 
data~\cite{H1_bbar,ZEUS_bbar,CDF_bbar,D0_bbar} ($K \sim 50$ for $J/\psi$
production and 
$K \sim 2-4$ for bottom
production at the TEVATRON), indicating that a significant
part of the cross section is still missing in the calculations. 
\par 
At small $x$ the structure function $F_2(x,Q^2)$ is proportional
to the sea quark density, which is driven 
by the gluon density. The standard QCD fits 
determine the parameters of
the initial parton distributions at a starting scale $Q_0$. With the help of the
DGLAP evolution equations these parton distributions are then evolved to any
other scale $Q^2$, with the splitting functions still truncated at fixed
${\cal O}(\alpha_s)$ (LO) or ${\cal O}(\alpha_s ^2)$ (NLO).
Any physics process in the fixed order scheme is then calculated via 
collinear factorization into the coefficient functions
$C^a(\frac{x}{z})$ and collinear (independent of $\kt$) 
parton density functions:
$f_a(z,Q^2)$:
\begin{equation}
\sigma = \sigma_0 \int \frac{dz}{z} C^a(\frac{x}{z}) f_a(z,Q^2)
\label{collinear-factorisation}
\end{equation}
At large energies (small $x$) the evolution of parton distributions proceeds over a large
region in rapidity $\Delta y \sim \log(1/x)$ and effects of finite transverse
momenta of the partons may become increasingly important.
Cross sections can then be $\kt$ - factorized~\cite{GLR,LRSS2,CCH,CE}
into an off-shell ($\kt$ dependent) partonic cross section
$\hat{\sigma}(\frac{x}{z},\kt^2) $
and a $\kt$ - unintegrated parton density function 
${\cal F}(z,\kt^2)$:
\begin{equation}
 \sigma  = \int 
\frac{dz}{z} d^2 \kt \hat{\sigma}(\frac{x}{z},\kt^2) {\cal F}(z,\kt^2)
\label{kt-factorisation}
\end{equation}
The unintegrated gluon density ${\cal F}(z,\kt^2)$ is 
described by the BFKL~\cite{\BFKL}
 evolution equation in the region of asymptotically large energies (small $x$). 
 An appropriate description valid for
both small and large $x$ is given by the CCFM evolution
equation~\cite{\CCFM}, resulting in an unintegrated gluon density 
${\cal A} (x,\kt^2,\Pmax^2 ) $, which is a function also of the 
additional scale $\Pmax $ described below.
Here and in the following we use the following
classification scheme: $x{\cal G}(x,\kt^2)$ describes DGLAP type unintegrated
gluon distributions, $x{\cal F}(x,\kt^2)$ is used for  pure BFKL and 
$x{\cal A}(x,\kt^2,\Pmax^2)$ stands for a CCFM type or any other type having two
scales involved.
\par
By explicitly carrying out the $\kt$ integration in eq.(\ref{kt-factorisation})
one can obtain a form fully consistent with collinear 
factorization~\cite{catani-feb2000,catani-dis96}: the
coefficient functions and also the DGLAP splitting functions leading to 
$f_a(z,Q^2)$ are no longer evaluated in fixed order perturbation theory but
supplemented with the all-loop resummation of the $\as \log 1/x$ contribution  
at small $x$. This all-loop resummation shows up in 
the \emph{Regge} form factor $\Delta_{Regge}$ for BFKL or in the
\emph{non-Sudakov} form factor $\Delta_{ns}$ for CCFM, 
which will be discussed in more 
detail in chapter~\ref{sec:un-integrated-parton}. 
\section{Off-shell matrix elements}
\label{sec:shell-matr-elem}
\begin{figure}[htb]
  \begin{center}
    \input{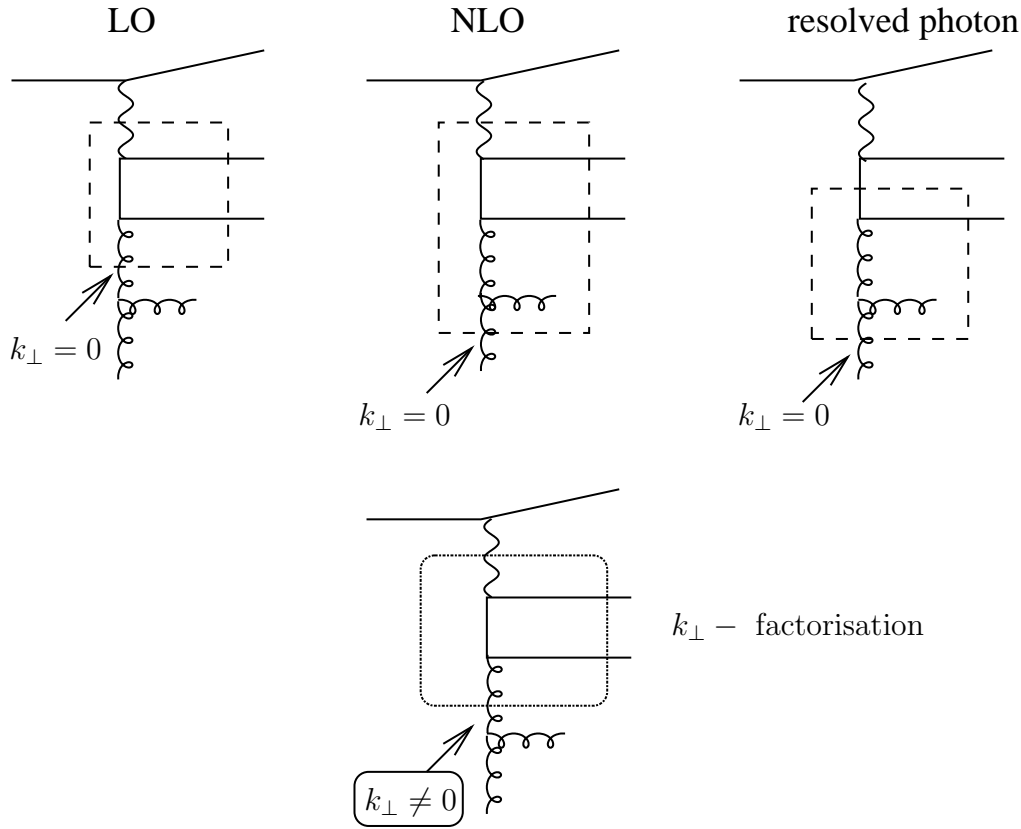}
    \caption{{\it Diagrammatic representation of LO, NLO and resolved photon
        processes in the collinear approach (top row) and compared to
        the \kt-factorization approach.}}
   \label{fig:nlo-kt-fact}
  \end{center}
\end{figure}
It is interesting to compare the basic features of the
\kt-factorization approach to the conventional collinear approach.  In
\kt-factorization the partons entering the hard scattering matrix
element are free to be off-mass shell, in contrast to the collinear
approach which treats all incoming partons massless.  The full advantage
of the \kt-factorization approach becomes visible, when additional
hard gluon radiation to a $ 2 \to 2$ process like $\gamma g \to
Q\bar{Q}$ is considered. If the transverse momentum \pti{g} of the
additional gluon is of the order of that of the quarks, then in a
conventional collinear approach the full \ord{\as^2} matrix element
for $2 \to 3$ has to be calculated.  In \kt-factorization such
processes are naturally included to leading logarithmic accuracy, even
if only the LO \as\ off-shell matrix element is used, since the \kt\ 
of the incoming gluon is only restricted by kinematics, and therefore can
aquire a virtuality similar to the ones in a complete fixed order
calculation.  In Fig.~\ref{fig:nlo-kt-fact} we show schematically the
basic ideas comparing the diagrammatic structure of the different
factorization approaches. Not only does \kt-factorization include (at
least some of the) NLO diagrams~\cite{RSS99}
 it also includes diagrams of the
resolved photon type, with the natural transition from real to virtual
photons.
\par
However, it has to be carefully investigated, which parts of a full
fixed NLO calculation are already included in the \kt-factorization
approach for a given off-shell matrix element and which are still
missing or only approximatively included. It should be clear from the
above, that the \ordas\ matrix element in \kt-factorization includes
fully the order \ordas\ matrix element of the collinear factorization
approach, but includes also higher order contributions. In addition,
due to the unintegrated gluon density, also parts of the virtual
corrections are properly resummed (Fig.~\ref{fig:kt-fact}b,c).

\subsection{Order \boldmath\as\ off-shell matrix elements} 
Several calculations exist for the process $\gamma g \to Q\bar{Q}$ and
$g g \to Q\bar{Q}$, where the gluon and the photon are both allowed to
be off-shell~\cite{\offshellmeqq} and $Q(\bar{Q})$ can be a heavy
or light quark (anti quark).
\begin{figure}[htb]
  \begin{center}
    \input{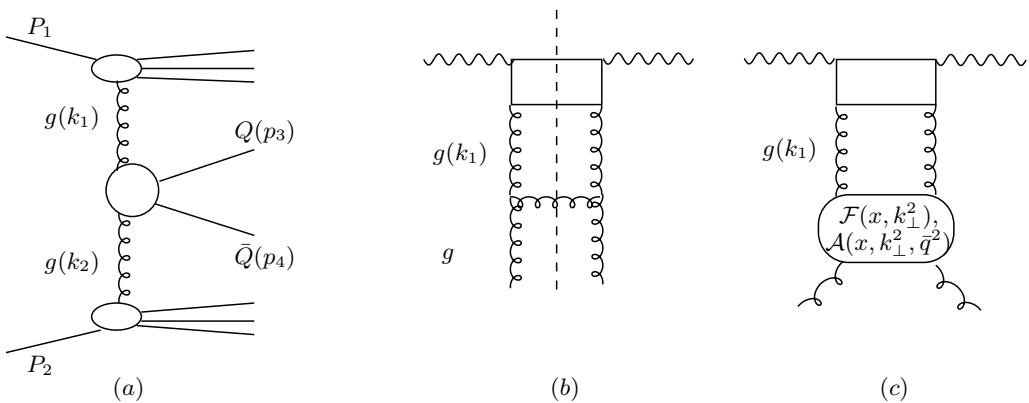}
    \caption{{\it Schematic diagrams for \kt-factorization: 
    $(a)$ shows the general case for hadroproduction of (heavy) quarks.
    $(b)$ shows the one-loop correction to the Born diagram for photoproduction
    $(c)$ shows the all-loop improved correction with the factorized structure
    function ${\cal F}(x,\kt^2)$ or ${\cal A}(x,\kt^2,\bar{q}^2)$.
    }}
   \label{fig:kt-fact}
  \end{center}
\end{figure}
\par
The four-vectors of the exchanged partons $k_1$, $k_2$ (in Sudakov
representation) are (Fig~\ref{fig:kt-fact}a):
\begin{eqnarray}
k_1^{\mu} =z_1 P_1^{\mu} + \bar{z}_1 P_2^{\mu} + \kti{1}^{\mu} \\
k_2^{\mu} =\bar{z}_2 P_1^{\mu} + {z}_2 P_2^{\mu} + \kti{2}^{\mu}  
\end{eqnarray} 
with $z_i,\bar{z}_i$ being the two components ($+,-$) of the light
cone energy fraction. The (heavy) quark momenta are denoted by $p_3$,
$p_4$ and the incoming particles (partons) by $P_1$, $P_2$ with
\begin{equation} 
P_{1,2} = \frac{1}{2}\sqrt{s}({\vec{0}},\pm 1,1), 
\hspace{0.5cm} 2P_1 P_2 =s
\end{equation}
where ${\vec{0}}$ indicates the vanishing two-dimensional
transverse momentum vector. 
In the case of photoproduction or leptoproduction this reduces to:
\begin{eqnarray}
k_1^{\mu} = P_1^{\mu}, \hspace{0.5cm} k_2^{\mu} = k^{\mu} \hspace{0.5cm}
\mbox{photo-production} \\
k_1^{\mu} = q^{\mu} = y P_1^{\mu} + \bar{y} P_2^{\mu} +\qti{2}^{\mu},
 \hspace{0.5cm} q^2 = -Q^2 \hspace{0.5cm} \mbox{leptoproduction}
\end{eqnarray} 
In all cases the off-shell matrix elements are calculated in the high
energy approximation, with $\bar{z}_1={z}_2=0$:
\begin{eqnarray}
k_1^{\mu} ={z}_1 P_1^{\mu} + \kti{1}^{\mu} \\
k_2^{\mu} =z_2 P_2^{\mu} + \kti{2}^{\mu},  
\end{eqnarray} 
which ensures that the virtualities are given by the transverse
momenta, $k_1^2 = -\kti{1}^2$ and $k_2^2 = -\kti{2}^2$.
The off-shell matrix elements involve 4-vector products not only with
$k_1,k_2$ and the outgoing (heavy) quark momenta $p_3,p_4$, but also
with the momenta of the incoming particles (partons) $P_1,P_2$. This
is a result of defining the (off-shell) gluon polarization tensors in
terms of the gluon ($k_i$) and the incoming particle vectors
($P_{1,2}$), which is necessary due to the off-shellness. In the
collinear limit, this reduces to the standard polarization tensors.
In~\cite{CCH,CE} it has been shown analytically, that the off-shell
matrix elements reduce to the standard ones in case of vanishing
transverse momenta of the incoming (exchanged) partons $k_1$, $k_2$.
\par
Due to the complicated structure of the off-shell matrix elements, it
is also necessary to check the positivity of the squared matrix
elements in case of incoming partons which are highly off shell 
($k_1^2,k_2^2 \ll 0$). It
has been proven analytically in \cite{Ciafaloni_me_2001} for the case
of heavy quarks with both incoming partons being off mass shell.
 We have also checked numerically,
that the squared matrix elements are positive everywhere in the phase
space, if the incoming particles (electron or proton) are exactly massless
($P_1^2=P_2^2=0$).  As
soon as finite masses (of the electron or proton)
are included, the exact cancellation of
different terms in the matrix elements is destroyed, and unphysical
(negative) results could appear.
\par
From the off-shell matrix elements for a $2 \to 2$ process, it is easy
to obtain the high energy limit of any on-shell $2 (\mbox{massless
partons}) \to 2 + n (\mbox{massless partons}) $ process, with
$n=1,2$ (see Fig.~\ref{fig:kt-fact}b for the case of photoproduction),
by applying a correction as given in \cite[page 180]{CCH}. By doing so,
the corrections coming from additional real gluon emission are estimated (i.e.
$\gamma g \to Q \bar{Q} g$ corrections to $\gamma g \to Q \bar{Q}$) with the
incoming gluons now treated on-shell. In the case of hadroproduction one obtains
the NNLO corrections $g g \to Q \bar{Q} g g$. In this sense 
\kt-factorization provides
an easy tool for estimating $K$-factors in the high energy limit:
\begin{equation}
   K = \frac{\sigma_{tot}}{\sigma(\mbox{LO or NLO})}
   \simeq\frac{\sigma(2 \to 2) + \sigma(2 \to 2 + n)}{\sigma(2 \to 2)}
\end{equation}
\par
In Tab.~\ref{table:offshell} we give an overview over the different
off-shell matrix elements available. It has been checked, that  
the different calculations \cite{CCH,saleev_zotov_b} of the process 
\mbox{$\gamma^{(*)} g^* \to Q\bar{Q}$} 
give numerically the same result. Also the matrix elements of 
\cite{CCH,CE,LiSaZo} are identical.
\begin{table}[htb]
\begin{center}
\begin{tabular}{| l | l | l |} 
\hline
process  & comments & reference\\
\hline
$\gamma g^* \to Q \bar{Q}$ & $|M|^2$ & 
\protect\cite{CCH,saleev_zotov_b,LipSalZot2000} \\
$\gamma^* g^* \to Q \bar{Q}$ & $|M|^2$ &  
\protect\cite{CCH,saleev_zotov_b} \\
$\gamma^* g^* \to Q \bar{Q}$ & partially integrated &  \protect\cite{Salam,KLPZ} \\
$g^* g^* \to Q \bar{Q}$ & $|M|^2$ &  \protect\cite{CCH,CE,LiSaZo} \\
\hline
$\gamma g^* \to J/\psi g$ & $|M|^2$ &\protect\cite{saleev_zotov_a,LiZo2000} \\
$\gamma g^* \to J/\psi g$ & helicity amplitude &\protect\cite{baranov} \\
$\gamma^* g^* \to J/\psi g$ & helicity amplitude & \protect\cite{baranov} \\
$g^* g^* \to J/\psi g$ & helicity amplitude & \protect\cite{baranov_gg-jpsig} \\
\hline
\end{tabular}
\caption{{\it Table of calculations of 
    \ordas\ off-shell matrix elements for different types of
    processes.  }}
\label{table:offshell}
\end{center}
\end{table}

\subsection{Next-to-Leading corrections to
 \boldmath\as\ off-shell matrix elements} 
 \label{susubsec:NLLme}
So far we have been dealing with matrix elements only in \ordas, but
even here \kt-factorization has proven to be a powerful tool in the
sense that it includes already in lowest order a large part of the NLO
corrections~\cite{RSS99}
 of the collinear approach.  However, LO calculations are
still subject to uncertainties in the scale of the coupling and in the
normalization. As the next-to-leading corrections (including
energy-momentum conservation) to the BFKL equation are now available
also the matrix elements need to be calculated in the next-to-leading
order.
\begin{figure}
  \begin{center}
    \input{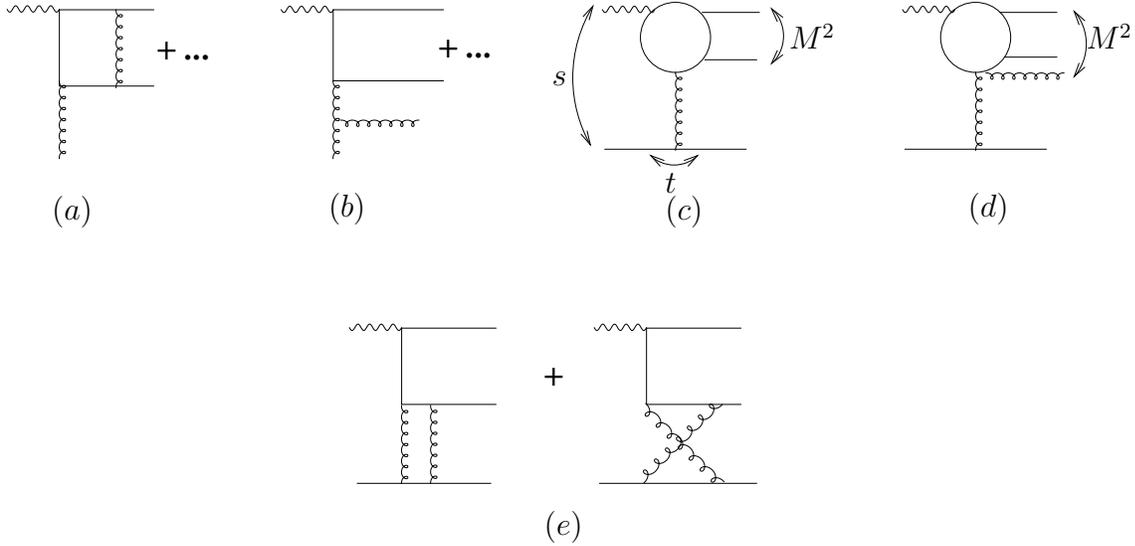}
    \caption{{\it Schematic diagrams for next-to-leading contributions:
    $(a)$ virtual corrections, $(b)$ real corrections, $(c)$ the process 
    $\gamma^* +q \to (q \bar{q})+q$,
    $(d)$ the process $\gamma^* +q \to (q \bar{q} g)+q$,
    $(e)$ diagrams showing the reggeization of the gluon
    }}   \label{fig:nlo}
  \end{center}
\end{figure}
\par
NLO corrections to the process $\gamma^*g^* \to X$ contain 
the virtual corrections to $\gamma^*g^* \to q \bar{q}$ (Fig.~\ref{fig:nlo}$a$) 
and the leading order off-shell matrix element for the process 
$\gamma^* g^* \to q \bar{q} g$ (Fig.~\ref{fig:nlo}$b$). The calculation of the 
former ones has been completed, and the results for the matrix elements 
are published in~\cite{bartels-nlo-qqbar-virtual}. 
The real corrections ($\gamma^* g^* \to q \bar{q} g$)
have been obtained for helicity-summed squares of the matrix elements. 
For the longitudinally polarized photon they are published 
in~\cite{bartels-nlo-qqbar-real_1}, 
and results for the transversely polarized photon will be made available 
soon~\cite{bartels-nlo-qqbar-real_2}.
\par 
The starting point of these calculations is a study of the Regge limit of the 
processes $\gamma^* +q \to (q \bar{q})+q$ (Fig.~\ref{fig:nlo}$c$) and 
$\gamma^* +q \to (q \bar{q} g)+q$ (Fig.~\ref{fig:nlo}$d$), i.e. 
the contributing QCD diagrams have been calculated in the region  
$s=(q+p)^2 \gg Q^2,\;M^2,\;t$ and $Q^2 \gg \Lambda_{QCD}$,
with $q (p)$ being the four-vector of the photon (quark). In this limit, the 
exchanged gluon is not an elementary but a reggeized gluon: the two gluon 
exchange diagrams (Fig.~\ref{fig:nlo}$e$) contain a term proportional to 
$\omega(t) \log (s)$, 
where $\omega(t)$ is the gluon Regge trajectory function. This term is not 
part 
of the subprocess $\gamma^*g^* \to q \bar{q}$ but rather belongs to the 
exchanged gluon. In order to find the 
contribution of Fig.~\ref{fig:nlo}$e$ to this scattering subprocess, 
one first has to 
subtract the reggeization piece. This means that, for the 
subprocess $\gamma^*g^* \to q \bar{q}$, the notation `$g^*$'
in NLO not only stands for `off-shellness' but also for `reggeized'. It 
also has important consequences for the 
factorization of the NLO off-shell subprocess inside a larger QCD diagram:
the $t$-channel gluons connecting the different subprocesses 
(see, for example, $g(k_1)$ in Fig.~\ref{fig:kt-fact}) are not elementary 
but reggeized,
and the QCD diagrams include subsets of graphs with more than two gluons 
in the $t$-channel (see Fig.~\ref{fig:nlo}$e$) 
\par  
The results for the virtual corrections 
contain infrared singularities, both soft and virtual ones. As usual, 
when integrating the real corrections over the final state gluon,
one finds the same infrared singularities but with opposite signs. So in the 
sum of virtual and real corrections one obtains an infrared finite answer.
A peculiarity of this NLO calculation is the interplay with the 
LO process $\gamma^* +q \to (q\bar{q}) + g +q$. 
In the latter, the process is calculated in the 
leading $\log (s)$ approximation (LO BFKL approximation),
 where the gluon has a large rapidity 
separation w.r.t. the $q\bar{q}$-pair 
(i.e. it is emitted in the central region between the $q\bar{q}$-pair and the
lower quark line in Fig.~\ref{fig:nlo}$d$). 
As a result of 
this high energy (small $x$- or Regge) factorization,
expressions for the vertex $g^* + g^* \to g$ and for the subprocess 
$\gamma^* g^* \to q\bar{q}$ are obtained. When turning to the NLO corrections 
of the process $\gamma^* +q \to (q \bar{q} g)+q$, the calculation 
extends to next-to-leading $\log (s)$ accuracy, but it contains, as a special 
case, also the LO BFKL process. In order to avoid double counting, 
one has to subtract the central region contribution.
Only after this subtraction 
we have a clean separation: $q\bar{q}g$ final states without or with a large 
rapidity separation between the gluon and the quark-antiquark pair. 
The former one 
belongs to the NLO approximation, whereas the latter one is counted in 
the leading $\log (s)$ approximation.
\par
The results for real corrections in~\cite{bartels-nlo-qqbar-real_1} 
are very interesting also in 
the context of the QCD color dipole picture. It is well-known that the total 
LO $\gamma^*q$ cross section at high energies can be 
written in the form~\cite{Mueller_1994,Nikolaev_1994}:
\begin{equation}
\sigma_{tot}^{\gamma^*q} = \int dz \int d \rho 
\left( \psi_{q\bar{q}}^{\gamma^*}(Q,z,\rho) \right)^* 
\sigma_{q\bar{q}} (x,z,\rho) 
\psi_{q\bar{q}}^{\gamma^*} (Q,z,\rho)
\label{color-dipole}
\end{equation}
where $z$ and $(1-z)$ denote the momentum fractions (w.r.t. the photon 
momentum $q$) of the quark and the antiquark, respectively,
 $\rho$ is the transverse 
extension of the quark-antiquark pair,
$\psi_{q\bar{q}}$ stands for the $q\bar{q}$ Fock component of the photon
wave function, and $\sigma_{q\bar{q}}$ is the color dipole 
cross section.
This form of the scattering cross section is in agreement with the space time
picture in the target quark rest frame: a long time before the photon 
reaches the 
target quark at rest, it splits into the quark-antiquark pair which
then interacts with the target. During the interaction the transverse 
extension of the quark-antiquark pair remains frozen, i.e the initial 
quark-antiquark pair has the same $\rho$-value as the final one. 
Via the optical theorem the 
$\gamma^*q$ total cross section is related to the square of the 
scattering matrix element of the process $\gamma^*+q \to (q\bar{q})+q$;
the dipole cross section form eq.(\ref{color-dipole})
 must then be a consequence of the special 
form of the LO amplitude for the subprocess $\gamma^* g^* \to q\bar{q}$.
Since this form of the total $\gamma^*q$ cross section (which is preserved 
when the target quark is replaced by the target proton) looks so 
appealing (and also has turned out to be extremely useful in phenomenological 
applications), 
that it is desirable to investigate its validity also beyond 
leading order.
\par
When trying to generalize 
eq.(\ref{color-dipole}) to NLO, one is led to study the square of 
the real corrections $\gamma^*+q \to (q\bar{q}g) +q$. If the color dipole 
picture remains correct (factorization in wave function and dipole cross
section), 
this contribution naturally should lead to a new 
$q\bar{q}g$ Fock component of the photon wave function, and to a new 
interaction
cross section $\sigma_{q\bar{q}g}$, which describes the interaction of the
quark-antiquark-gluon system with the target quark. 
In~\cite{bartels-nlo-qqbar-real_1,bartels-nlo-qqbar-real_2} it 
is shown that this is indeed the case. The form of the new photon wave 
function 
is rather lengthy (in particular for the transverse 
photon), and a more detailed investigation is still needed. Nevertheless,  
one feels tempted to conclude that 
the color dipole form eq.(\ref{color-dipole})
is the beginning of a general color multipole 
expansion, where the different Fock components of the photon wave function 
describe the spatial distribution of color charge inside the photon. 
However, before this conclusion can really be drawn, it remains to be shown 
that also the virtual corrections to the scattering amplitude of 
$\gamma^* +q \to (q\bar{q}) +q$ fit into the form eq.(\ref{color-dipole}). 
These corrections 
should lead to NLO corrections of the photon wave function or the dipole 
cross section. They also may slightly `melt' the transverse extension of the 
quark-antiquark pair during the interaction with the target.
\par
In summary, the results 
in~\cite{bartels-nlo-qqbar-virtual,bartels-nlo-qqbar-real_1,%
bartels-nlo-qqbar-real_2} provide the starting point 
for discussing exclusive final states in the $\kt$-factorization scheme
 in NLO accuracy.
However, before these formulae can be used in a numerical analysis, 
virtual and real corrections have to be put together, and IR finite 
cross sections have to be formulated. The NLO corrections to the off-shell
matrix elements described in this subsection also represent the
main ingredients to the NLO photon impact factor. Its importance
will be discussed in section \ref{susubsec:NLLinvestigations}    
\subsection{Gauge-invariance}  
The off-shell matrix elements and the cross section taken in lowest order
$\alpha_s$ are gauge invariant, as argued in~\cite{CCH}. However, when extended
to next order in perturbation theory, as depicted in 
Fig.~\ref{fig:nlo-kt-fact}, problems will occur and the 
off-shell matrix elements and unintegrated parton distributions are no longer
necessarily
gauge invariant.  The following critique indicates  
that their definition needs further specification and that further work is
needed to properly justify the formalism~\cite{Collins-lund-smallx}.
\par
Basically, parton distributions are defined as a hadron expectation value 
of a quark and antiquark field (or two gluon fields): 
\begin{equation} 
  \langle p| \bar\psi(y) \Gamma \psi(0) |p\rangle, 
\end{equation} 
with an appropriate Fourier transformation on the space-time argument 
$y^\mu$, some appropriate Dirac matrix $\Gamma$ and some appropriate 
normalization.  This definition is not gauge invariant, so one must 
either specify the gauge or one modifies the definition to make it 
gauge invariant: 
\begin{equation} 
  \langle p| \bar\psi(y) \Gamma P e^{-ig\int_0^y dy'^\mu 
  A^\alpha_\mu(y')t_\alpha} \psi(0) |p\rangle. 
\end{equation} 
Here, $t_\alpha$ are generating matrices for the $SU(3)$ color group of 
QCD, and the symbol $P$ denotes that the gluon fields are laid out in 
their order on the path in the integral from $0$ to $y$.   
\par
The big question is which path is to be used. 
For the integrated parton distributions, $y$ is at a light-like separation  
from $0$: normally $y^- \neq 0$, $y^+=y_\perp=0$, and then taking the path 
along the light-like straight line joining $0$ and $y=(0,y^-,0_\perp)$ is  
natural and correct.  But for unintegrated distributions, this is not so 
simple; one has a choice of paths.
The choice is not arbitrary but is determined by the
derivation: i.e., the definition is whatever is appropriate to make a
correct factorization theorem.  
\par
This can be seen from the analysis of gluon emission that we have
already described.  This analysis is only applicable if ladder diagrams,
as in Fig.~\ref{fig:nlo-kt-fact},
actually dominate.  In fact, in a general gauge,
non-ladder diagrams are as important as ladder diagrams.  This is the
case, for example, in the Feynman gauge.  The standard
leading-logarithm analysis suggests, as is commonly asserted, that the
appropriate gauge is a light-cone gauge $n \cdot A = 0$, where $n$ is a
light-like vector in a suitable direction, for then non-ladder
contributions are suppressed in the LLA analysis.  From this one would
conclude that the off-shell matrix elements and the unintegrated
parton distributions are defined as the appropriate field theory Green
functions in light-cone gauge.
\par
Closer examination shows that there must be problems since the
unintegrated parton distributions are divergent.  This was shown quite
generally by Collins and Soper~\cite{CS1,CS2}
as part of their derivation
of factorization for transverse-momentum distributions.
They found they needed to derive factorization and to define
unintegrated parton distributions in a {\em non}-light-like axial gauge,
i.e., with $n^2 \neq 0$.\footnote{ An 
  equivalent and probably better definition can be made with suitable
  path-ordered exponential in the operator.  } 
They derived an evolution equation for the dependence on the gauge-fixing
vector; the evolution is quite important and contains doubly-logarithmic
Sudakov effects.  In the limit $n^2 \to 0$, the parton distributions become so
singular that they are not defined.  The divergences are associated with
the emission of gluons of arbitrarily negative rapidity with respect to
the parent hadron; commonly these are called soft gluons, and a
non-light-like gauge-fixing vector provides a cutoff on the divergences. 
It is normally said that soft gluons cancel in QCD cross sections. 
However, the standard cancellation requires an integral over all parton 
transverse momentum, which is of course not applicable in an unintegrated 
parton density. 
\par
The divergences can readily be seen in one-loop graphs, and are a
generalization of divergences known to occur in light-cone gauge.
Modification of the $i\epsilon$ prescription of the $1/ k \cdot n$
singularity of the gluon propagator, as is commonly advocated, is not
sufficient, since this does not remove the divergence in the emission of
{\em real} gluons, for which the singularity is an endpoint singularity in
$k \cdot n$.
\par
Clearly, in all the derivations of the BFKL and related equations, there
must be an appropriate cutoff.  Unfortunately, this is not normally made
very explicit, even though an explicit specification of the cutoff is
vitally important if the formalism is to make sense beyond LO. 
One can expect, that such a cutoff is a cutoff in angle or rapidity,
and it should be related to angular ordering, supporting the
intuitive approximate picture. However, the
presence of the cutoff implies that there is an extra parameter in the
parton distributions, whose variation should be understood.  Balitsky
\cite{Balitsky} has worked in this direction, although his formulation
does not appear to have developed far enough to address the issues in the
present document.
\par
A proper derivation will also result in an explicit definition of a
reggeized gluon.  Such a definition is not readily extracted from the
original BFKL publications.  At most these publications provide a
definition as a property of a solution of their equation.  No explicit
definition is given whereby quantities involving reggeized gluons are
matrix elements of some operator or 
other.

\section{Unintegrated parton distributions}
\label{sec:un-integrated-parton}
\subsection{Introduction}
The conventional parton distributions describe the density of partons
carrying a certain longitudinal momentum fraction inside the proton.
These distributions are integrated over the transverse momenta of the
partons up to the factorization scale $\mu$. However, in order to
describe some exclusive processes it becomes necessary to consider the
transverse momenta of the partons and thus use so called unintegrated
gluon distributions ${\cal A}(x,\kt^2,\mu^2)$. 
Unintegrated parton distributions account for the 
resummation of a variety of logarithmically enhanced terms, such as
\begin{equation}
\left(\as \log(\mu^2/\Lambda^2)\right)^n,  
\left(\as\log(\mu^2/\Lambda^2)\log(1/x)\right)^n,
\left(\as \log(1/x)\right)^n
 \mbox{ and } 
\left(\as \log^2(\mu^2/\kt^2)\right)^n
\label{large_logs}
\end{equation} 
The
unintegrated parton distributions describe the probability of
finding a parton carrying a longitudinal momentum fraction $x$ and a
transverse momentum \kt\ at the factorization scale $\mu$.  The
unintegrated gluon density $x{\cal A}(x,\kt^2,\mu^2)$ can be related to
the integrated one $xg(x,\mu^2)$ by:
\begin{equation}
xg(x,\mu^2) \simeq \int_0^{\mu^2} d\kt^2 x{\cal A}(x,\kt^2,\mu^2)
\label{intglu}
\end{equation}
The $\simeq$ sign in the above equation indicates, that there is no
strict equality between unintegrated and integrated parton
distributions, as neither are observables.
\par
In unintegrated parton distributions, the contribution from the large
logarithms in $x$ and $\mu^2$, the terms specified above in expression
(\ref{large_logs}),
 can be correctly disentangled.  However, the unintegrated parton
distributions are defined only as a functions of three variables $x$,
$\kt^2$, $\mu^2$.  The 4-vector $k$ of the propagator gluon in initial
state cascade is given by:
\begin{displaymath}
  k = (k^+, k^-, \kt)
\end{displaymath}
with $k^+ = x^+P^+$, $k^-=x^-P^-$  The virtuality $k^2$ is:
\begin{displaymath}
  k^2 = k^+k^- - \kt^2 = x^+x^-P^-P^+ - \kt^2.
\end{displaymath}
In the region of strongly ordered $k^+$, the typical values of $k^-$ are
small enough that it can be neglected in the hard scattering factor.  For
example, the virtuality is dominated by the transverse momentum only:
\begin{displaymath}
  k^2 \simeq -k_\perp^2.
\end{displaymath}
However, the $k^-$ integral is still present, and in fact, part of the
definition of an unintegrated parton density is, that the relevant
matrix element of a gluonic operator is integrated over all $k^-$. 
This is the reason, why only $x=x^+$ is kept in the unintegrated gluon density. 
\par
The all-loop resummation of the $\as \log 1/x$ contribution  
at small $x$ leads to  
\emph{Reggeization} of the gluon vertex, 
giving rise to a significant form factor. 
The \emph{Regge} form factor $\Delta\sub{Regge}$, often also called  
 \emph{non-Sudakov} form factor $\Delta\sub{ns}$,
regularizes the $1/z$ divergence in the splitting function,
\begin{equation}
  \label{eq:bfklsplit}
  P_g(z_+)\propto\frac{\Delta\sub{Regge}(\kt,z_+)}{z_+}.
\end{equation}
The $\Delta\sub{Regge}$ of BFKL is given by~\cite{Martin_Sutton2}:
\begin{equation}
\Delta\sub{Regge}(z_+,\kt^2)
 = \exp{\left(-\asb \int_{z_+}^1 \frac{dz'}{z'} 
\int \frac{dq_{\perp}^{'2}}{q_{\perp}^{'2}} \Theta(\kt^2 - q_{\perp}^{'2})  
\Theta(q_{\perp}^{'2}-q_0^2) \right)} 
\end{equation}
with $\alphasb=\frac{C_A \as}{\pi}=\frac{3 \as}{\pi}$ and $q_0$ being a lower
cutoff.
This form factor can be expanded by a power series and then
symbolically represented as:\\
\begin{tabular}{l l l l}
& 
\begin{minipage}[t]{0.15\textwidth}
\vspace*{-0.1cm} 
\rotatebox{0.}{\scalebox{0.4}{\includegraphics*{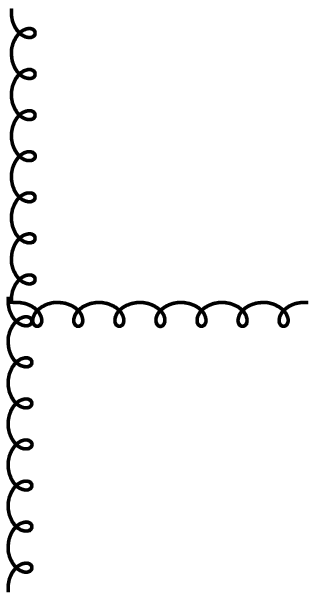}}}
\end{minipage}
& 
\begin{minipage}[t]{0.15\textwidth}
\vspace*{-0.1cm} 
\rotatebox{0.}{\scalebox{0.4}{\includegraphics*{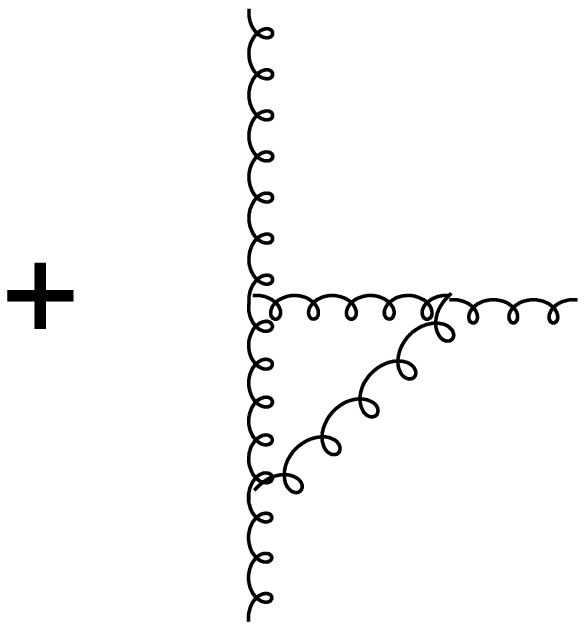}}}
\end{minipage}
& 
\begin{minipage}[t]{0.45\textwidth}
\vspace*{-0.1cm} 
\rotatebox{0.}{\scalebox{0.4}{\includegraphics*{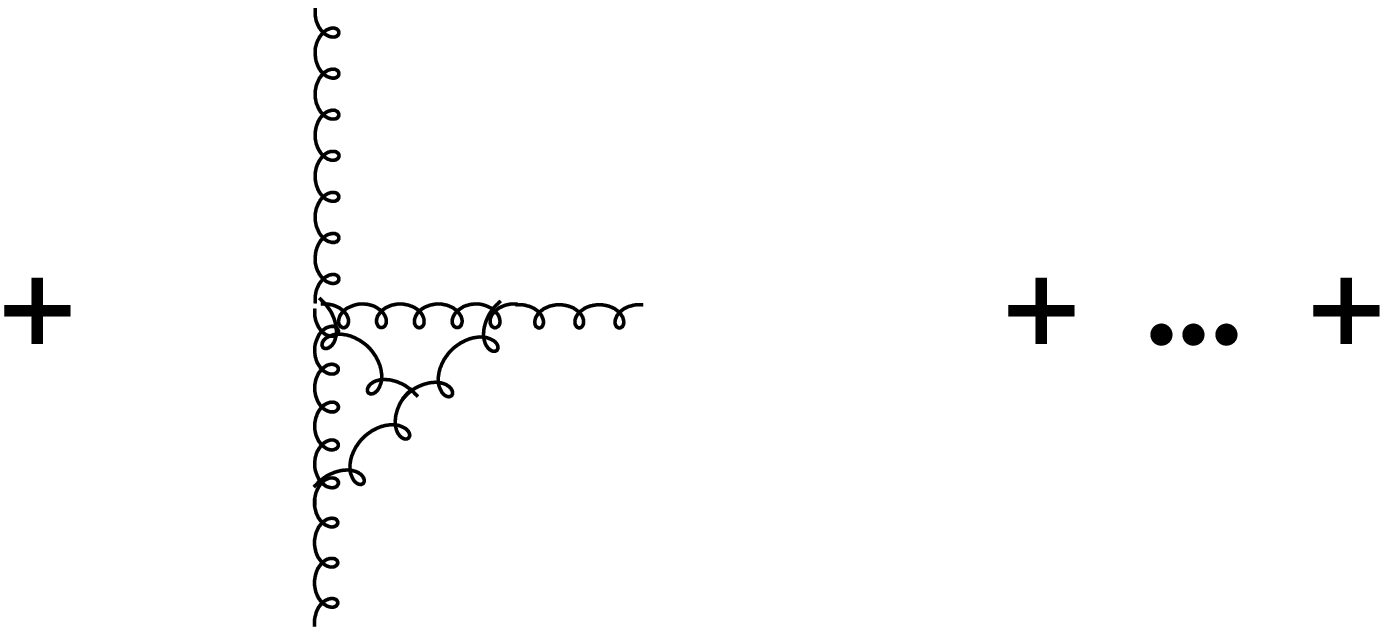}}}
\end{minipage}
\\
& $ \asb(\kt) \frac{1}{z_+}$ {\huge $\left[ \right.$} $1 $ & 
$+ \;\;\;\asb\log\left(z\right)
\log\left(\frac{\kt^2}{q^2_0}\right)
$ & $ + \;\;\; \left(\frac{1}{2!}\asb
\log\left(z\right)
\log\left(\frac{\kt^2}{q^2_0}\right)\right)^2  \;...$
{\huge $\left. \right]$} 
\end{tabular}
\vspace*{0.8cm}\\
where the small $x$ resummation of the virtual corrections becomes obvious.
Such a diagrammatic representation is of course gauge dependent.
It should be noted that the Regge form factor in this way looks like
the Sudakov form factor~\cite{DokshitzerQCD,EllisQCD} 
 used to regularize the $1/(1-z)$ pole in
standard DGLAP evolution,
 but the resummed diagrams involve small $z$
rather than large ones. This similarity is used in the derivation of
the Linked Dipole Chain model below.
\par
In interpreting the results quoted in this section, the reader 
should bear in mind the caveats explained in Sec.\ 3.3, that as yet,  
no fully explicit definition of the parton distributions has been 
given.  
\subsection{The CCFM evolution}
\label{sec:CCFM_evolution}
According to the CCFM~\cite{\CCFM} evolution equation the emission of
gluons during the initial cascade is only allowed in an
angular-ordered region of phase space.
The maximum allowed angle $\Xi$
is defined by the hard scattering quark box, producing the (heavy) quark pair.
 In terms
of Sudakov variables the quark pair momentum is written as:
\begin{equation}
p_q + p_{\bar{q}} = \Upsilon (P_1 + \Xi P_2) + \vec{Q}_{\perp}
\label{CCFMa}
\end{equation} 
where $\Upsilon$ ($\Upsilon  \Xi$) 
is the positive (negative) light-cone momentum fraction of the quark pair,
$P_{1,2}$ are the four-vectors of
 incoming beam particles, 
respectively and $\vec{Q}_t$ is the vectorial sum of the 
transverse momenta of the quark pair
in the laboratory frame.
Similarly, the momenta $p_i$ of the gluons emitted during the initial
state cascade are given by (here treated massless):
\begin{equation}
p_i = \upsilon_i (P_1 + \xi_i P_2) + p_{\perp i} \;  , \;\; 
\xi_i=\frac{p_{\perp i}^2}{s \upsilon_i^2},
\label{ccfm_q}
\end{equation}
with $\upsilon_i = (1 - z_i) x_{i-1}$, $x_i = z_i x_{i-1}$ and
$s=(P_1 + P_2)^2$ being the squared center of mass energy.
The variable $\xi_i$ is connected to the angle of the emitted gluon
with respect to the incoming proton and $x_i$ and $\upsilon_i$ are the
momentum fractions of the exchanged and emitted gluons, while $z_i$ is
the ratio of the energy fractions in the branching
$(i-1) \to i$ and $p_{\perp i}$ is
the transverse momentum of the emitted gluon $i$.
\par
The angular-ordered region is then specified by (Fig.~\ref{fig:variables}):
\begin{equation}
\xi_0 < \xi_1< \cdots < \xi_n < \Xi
\end{equation}
which becomes:
\begin{equation}
z_{i-1} \bar{q}_{i-1} < \bar{q}_{i} 
\label{ang_ord}
\end{equation}
where the rescaled transverse momenta $\bar{q}_{i}$ of the emitted
gluons is defined by:
\begin{equation}
 \bar{q}_{i} = x_{i-1}\sqrt{s \xi_i} = \frac{p_{\perp i}}{1-z_i}
\end{equation}
It is interesting to note, that the angular ordering constraint, as given by
eq.(\ref{ang_ord}), reduces to ordering in transverse momenta 
$p_{\perp}$ for large $z$,
whereas for $z\to 0$, the transverse momenta are free to perform a 
so-called random walk.  
\par
The CCFM evolution equation   with respect to the scale 
$\Pmax^2$ 
can be written in a differential
form~\cite{CCFMd}:
\begin{equation}
\Pmax^2\frac{d\; }{d \Pmax^2} 
   \frac{x \cA\left(x,\kt^2,\Pmax^2\right)}{\Delta_s(\Pmax^2,Q_0^2)}=
   \int dz \frac{d\phi}{2\pi}\,
   \frac{\tilde{P} \left(z,(\Pmax/z)^2,\kt^2\right)}{\Delta_s(\Pmax^2,Q_0^2)}\,
 x'\cA \left( x',\kt^{' 2},(\Pmax/z)^2\right) 
\label{CCFM_differential}
\end{equation} 
where $\cA(x,\kt^2,\Pmax^2)$ is the unintegrated gluon density, depending
on $x$, $\kt^2$ and the evolution variable $\mu^2= \Pmax^2$. The splitting
variable is $z=x/x'$ and $\vec{\kt}' = (1-z)/z\vec{q} + \vec{\kt}$,
where the vector $\vec{q}$ is at an azimuthal angle $\phi$.  The
Sudakov form factor $\Delta_s$ is given by:
\begin{equation}
\Delta_s(\Pmax^2,Q_0^2) =\exp{\left(
 - \int_{Q_0^2} ^{\Pmax^2}
 \frac{d q^{2}}{q^{2}} 
 \int_0^{1-Q_0/q} dz \frac{\alphasb\left(q^2(1-z)^2\right)}{1-z}
  \right)}
  \label{Sudakov}
\end{equation}
with $\alphasb=\frac{C_A \as}{\pi}=\frac{3 \as}{\pi}$. For
inclusive quantities at leading-logarithmic order the Sudakov form
factor cancels against the $1/(1-z)$ collinear singularity of the
splitting function.
\par
The splitting function $\tilde{P}$ for branching $i$ 
is given by:
\begin{equation}
\tilde{P}_g(z_i,q_i^2,\kti{i}^2)= \frac{\alphasb(q^2_{i}(1-z_i)^2)}{1-z_i} + 
\frac{\alphasb(\kti{i}^2)}{z_i} \Delta_{ns}(z_i,q^2_{i},\kti{i}^2)
\label{Pgg}
\end{equation}
where the non-Sudakov form factor $\Delta_{ns}$ is defined as:
\begin{equation}
\log\Delta_{ns}(z_i,q_i^2,\kti{i}^2) =  -\alphasb
                  \int_{z_i}^1 \frac{dz'}{z'} 
                        \int \frac{d q^2}{q^2} 
              \Theta(\kti{i}-q)\Theta(q-z'q_i)
                  \label{non_sudakov}                   
\end{equation}
The upper limit of the $z'$ integral is constrained by the $\Theta$
functions in eq.(\ref{non_sudakov}) by: $z_i \leq z^{\prime} \leq
\mbox{min}(1,\kti{i}/q_{i}) $, which results in the following form
of the non-Sudakov form factor \cite{Martin_Sutton}:
\begin{equation}
\log\Delta_{ns} = -\alphasb(\kti{i}^2)
\log\left(\frac{z_0}{z_i}\right)
\log\left(\frac{\kti{i}^2}{z_0z_i q_{i}^2}\right)
\label{ns_new}
\end{equation} 
where
$$z_0 = \left\{ \begin{array}{ll}
    1             & \mbox{if  } \kti{i}/q_{i} > 1 \\
    \kti{i}/\pti{i} & \mbox{if  } z_i < \kti{i}/q_{i} \leq 1 \\
    z_i             & \mbox{if  } \kti{i}/q_{i} \leq z_i  
  \end{array} \right. $$
\par
The unintegrated gluon density
$\cA(x,\kt^2,\Pmax^2)$ is a function also of the
angular variable $\Pmax$, ultimately limited by an angle, 
$\Pmax^2 =x^2_{n-1} {\Xi s}$, defined
by the hard interaction, and the 
two scales $\kt^2,\Pmax^2$ in  $\cA(x,\kt^2,\Pmax^2)$
should not come as a surprise. As we are
concentrating on an evolution, which is not ordered in transverse momenta,
it is natural to have more than one scale in the problem. The typical
example of such a two scale evolution process is
$\gamma^*\gamma^*\rightarrow\mbox{hadrons}$ scattering 
where the photons are highly virtual, but the virtualities still being
much smaller than the total energy, $s\gg
Q_1^2\approx Q_2^2 \gg \Lambda\sub{QCD}$. Another example is 
DIS with a forward high-\pt\ jet, where $Q^2$ and 
$\pt^2$ provide the scales. In the DGLAP approximation the
evolution is performed between 
 a small and a large scale -- a large scale probe
resolves a target at a smaller scale. It is obvious, that this evolution is not
appropriate for the case of two similar scales and   
instead an evolution in
rapidity, or angle, is needed.
\par 
In CCFM the scale $\Pmax$ (coming from the maximum angle) 
can be related to the
evolution scale in the collinear parton distributions.
This becomes obvious since 
\begin{equation}
\Pmax^2 \sim x_g^2 \Xi s= y x_g s 
 = \hat{s} + Q_{\perp}^2
\end{equation}
The last expression is derived by using 
$p_q + p_{\bar{q}} \simeq x_g P_2 + y P_1 + \vec{Q}_{\perp}$,
$\Xi  \simeq y/x_g$ and $\hat{s} = y x_g s - Q_{\perp}^2$.
This can be compared to a possible choice of 
the renormalization and factorization scale $\mu^2$  
in the collinear approach with $\mu^2 = Q_{\perp}^2 + 4 \cdot m_q^2$ and the
similarity between $\mu$ and $\Pmax$ becomes obvious.

\subsection{LDC and the transition between BFKL and DGLAP}
\label{sec:trans-betw-bfkl}

The Linked Dipole Chain (LDC) model~\cite{LDCa,LDCb}
 is a reformulation of CCFM which
makes the evolution explicitly left--right symmetric, which will be discussed in
section~\ref{subsec:beyond-leading-order} in more detail. LDC relies on
the observation that the non-Sudakov form factor in  
eq.(\ref{non_sudakov}), despite its name, can be interpreted as a kind of
Sudakov form factor giving the no-emission probability in the region
of integration. 
To do this an additional constraint on the initial-state
radiation is added requiring the transverse momentum of the
emitted gluon to be above the smaller of the transverse momenta of the
connecting propagating gluons:
\begin{equation}
  \label{eq:ldccut}
  \pti{i}^2>\min(\kti{i-1}^2,\kti{i}^2).
\end{equation}
Emissions failing this cut will instead be treated as final-state
emissions and need to be resummed in order not to change the total
cross section. In the limit of small $z$ and imposing the kinematic 
constraint of eq.(\ref{eq:CC}) (see section~\ref{subsec:beyond-leading-order}
for further details) 
gives a factor which if it is multiplied with
each emission, completely cancels the non-Sudakov and thus, in a sense
\textit{de-reggeizes} the gluon.

One may ask when it is appropriate to use collinear factorization
(DGLAP) and when we must account for effects of BFKL and
\kt-factorization. Clearly, when \kt\ is large and $1/x$ is limited we
are in the DGLAP regime, and when $1/x$ is large and \kt\ is limited
we are in the BFKL regime. An essential question is then: What is
large?  Where is the boundary between the regimes, and what is the
behavior in the transition region? Due to the relative simplicity of
the LDC model, which interpolates smoothly between the regimes of
large and small \kt, it is possible to give an intuitive picture of
the transition. In the following $x$ is always kept small, and leading
terms in $\log 1/x$ are studied. Thus the limit for large \kt\ does not
really correspond to the DGLAP regime, but more correctly to the
double-log approximation.
\par
First we will discuss the case of a fixed coupling. For large values
of \kt\ (in the ``DGLAP region'') the unintegrated structure function
${\cal F}(x,\kt^2)$ is dominated by contributions from chains of gluon
propagators which are ordered in \kt.  The result is a product of
factors $ \alb \cdot \frac{dx_i }{ x_i} \cdot \frac{d\kti{i}^2
  }{\kti{i}^2}$ \cite{DGLAPa,DGLAPb,DGLAPc,DGLAPd}:
\begin{eqnarray}
 {\cal F}(x,\kt^2) &\sim& \sum_N \int \prod^N \alb \,\, 
 dl_i \theta(l_{i+1} -l_i)\,\, d\kappa_i \Theta(\kappa_{i+1} - \kappa_i) 
 \nonumber \\
{\mathrm {where}} \,\,\,\, \alb\!\!\!\! &\equiv &\!\!\!\! \frac{3\as}{\pi}, 
\,\,\,\,l \equiv \log(1/x)\,\,\,\,{\mathrm {and}} 
\,\,\,\,\kappa \equiv \log(\kt^2/\Lambda^2)
\label{dglap}
\end{eqnarray}
Integration over $\kappa_i$ with the $\Theta$-functions gives the
phase space for $N$ ordered values $\kappa_i$. The result is
$\kappa^N/N!$. The integrations over $l_i$ give a similar result, and
thus we obtain the well-known double-log result
\begin{equation}
 {\cal F}(x,k_\perp^2) \sim \sum_N \bar{\alpha}^N \cdot \frac{l^N}{N!} \cdot \frac{\kappa^N}{N!} = I_0(2\sqrt{\bar{\alpha} l \kappa})
\label{DGLAP}
\end{equation}
In the BFKL region also non-ordered chains contribute, and the result
is a power-like increase $\sim x^{-\lambda}$ for small $x$-values
\cite{BFKLa,BFKLb,BFKLc}.
\par
The CCFM \cite{CCFMa,CCFMb,CCFMc,CCFMd} and the LDC model
\cite{LDCa,LDCb,LDCe} interpolate between the DGLAP and BFKL regimes.
In the LDC model the possibility to ``go down'' in $k_\perp$, from
$\kappa'$ to a smaller value $\kappa$, is suppressed by a factor
$\exp(\kappa - \kappa')$.  The effective allowed distance for downward
steps is therefore limited to about one unit in $\kappa$. If this
quantity is called $\delta$, the result is, that the phase space
factor $\frac{\kappa^N}{N!}$ is replaced by
$\frac{(\kappa+N\delta)^N}{N!}$. 
\\
Thus we obtain:
\begin{equation}
 {\cal F}(x,k_\perp^2) \sim \sum_N \frac{(\bar{\alpha} l)^N}{N!} 
 \frac{(\kappa+N\delta)^N}{N!}
\label{fixsumma}
\end{equation}
When $\kappa$ is very large, this expression approaches 
eq.(\ref{DGLAP}). When $\kappa$ is small we find, using Sterling's
formula, that ${\cal F}(x,\kt^2) \sim \sum_N (\alb l \delta e)^N / N!
= \exp (\lambda l) \equiv x^{-\lambda}$, with $\lambda = \alb \delta
e$.  For $\delta = 1$ this gives $\lambda = e \alb = 2.72\alb$, which
is very close to the leading order result for the BFKL equation,
$\lambda = 4 \log 2 \, \alb = 2.77\alb$.  Thus eq.  (\ref{fixsumma})
interpolates smoothly between the DGLAP and BFKL regimes. It is also
possible to see that the transition occurs for a fixed ratio between
$\kappa$ and $l$, $\kappa / l \approx e \alb$.
\par
For a running coupling a factor $\alb\, d\kappa$ in eq.(\ref{dglap})
is replaced by $\alpha_0\, d \kappa/\kappa = \alpha_0\, du$, where
$\alpha_0 \equiv \alb \kappa$, $u\equiv \log(\kappa/\kappa_0)$ and
$\kappa_0$ is a cutoff.  In the large \kt\ region (the ``DGLAP
region'') the result is therefore similar to eq. (\ref{DGLAP}), but
with $\alb\kappa$ replaced by $\alpha_0u$.  For
small $x$-values we note that the allowed effective size of downward
steps, which is still one unit in $\kappa$, is a sizeable interval in
$u$ for small $\kappa$, but a very small interval in $u$ for larger
$\kappa$.
\par
This is the reason for an earlier experience \cite{LDCb}, which
showed that a typical chain contains two parts. In the first part the
$k_\perp$-values are relatively small, and it is therefore easy to go
up and down in $k_{\perp}$, and non-ordered $k_\perp$-values are
important. The second part is an ordered (DGLAP-type) chain, where
$k_\perp$ increases towards its final value.
\par
\begin{figure}
  \begin{center}
    \epsfig{file=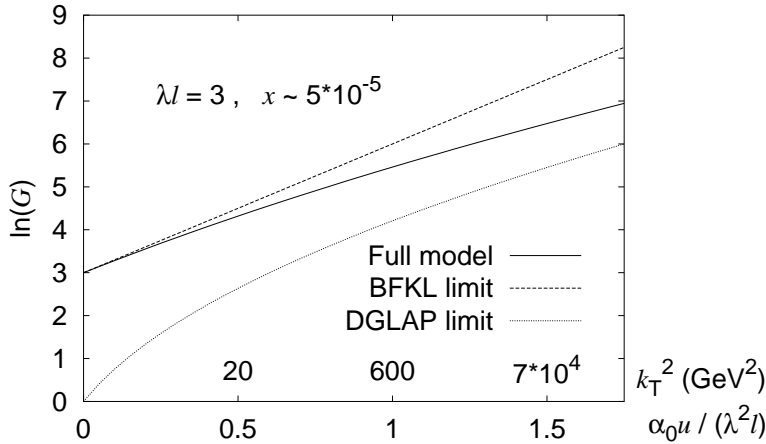,width=9cm}
    \caption[dummy]{{\it Results for a running coupling for $\lambda l = 3$,
      which for $\lambda=0.3$ means $x=5\cdot 10^{-5}$. (Note that this
      $x$-value corresponds only to the chain between the ``observed''
      low energy gluon and the parent soft gluon.) The solid line
      corresponds to the model in eq.  (\ref{Gslut}), the dashed line
      to the BFKL limit, eq.  (\ref{bfklrunning}), and the dotted line
      to the DGLAP limit in eq.  (\ref{dglaprunning}). The abscissa is
      the variable $w=\alpha_0 u/\lambda^2 l$ defined so that the
      transition occurs for $w\approx 1$.  The corresponding values
      for $\kt^2$ in $GeV^2$ for $\lambda=0.3$ and $\lambda l = 3$
      are also indicated.}}
    \label{fig:transition}
  \end{center}
\end{figure}

Let us study a chain with $N$ links, out of which the first
$N-k$ links correspond to
the first part with small non-ordered $k_\perp$ values,
 and the remaining $k$
links belong to the second part with increasing $k_\perp$. Assume that
the effective phase space for each $u_i$ in the soft part is given by
a quantity $\Delta$. Then the total weight for this part becomes
$\Delta^{N-k}$. For the $k$ links in the second, ordered, part the
phase space becomes as in the DGLAP case $u^k/k!$. Thus the total
result is ($G \equiv \kappa {\cal F}$)
\begin{equation}
G \sim \sum_N \frac{(\alpha_0 l)^N}{N!} \sum_{k=0}^N \frac{u^k}{k!} 
\Delta^{N-k} 
=\sum_N \frac{(\alpha_0 l \Delta)^N}{N!} \sum_{k=0}^N \frac{(u/\Delta)^k}{k!}
\,\,.
\label{Gslut}
\end{equation}
This simple model also interpolates smoothly between the DGLAP and
BFKL regions. For large $u$-values the last term in the sum over $k$
dominates, which gives the ``DGLAP'' result
\begin{equation}
G \sim \sum (\alpha_0 l u)^N/(N!)^2 = I_0(2\sqrt{\alpha_0 l u}) \approx 
(16 \pi^2 \alpha_0 l u)^{-1/4} \cdot \exp(2\sqrt{\alpha_0 l u}).
\label{dglaprunning}
\end{equation}
For small $u$-values the sum over $k$ gives approximately 
$\exp \left(u/\Delta\right)$, and thus
\begin{equation}
G \sim \exp (\alpha_0 \Delta l) \cdot \exp(u/\Delta) =x^{-\lambda} 
\kappa^{\alpha_0/\lambda}, \,\,\mathrm{with}\,\,\, \lambda=\alpha_0 \Delta.
\label{bfklrunning}
\end{equation}
The transition between the regimes occurs now for a fixed ratio
between $u / l \approx \lambda ^2 / \alpha_0$, which is of order 0.1
if $\lambda$ is around 0.3. The result is illustrated in fig.
\ref{fig:transition}, which shows how the expression in eq. (\ref{Gslut})
interpolates between the DGLAP and BFKL limits in eqs.
(\ref{dglaprunning}) and (\ref{bfklrunning}) for large and small
$k_\perp$.\footnote{
In the framework of $N=4$ SUSY, where the corresponding coupling 
is not running, the relations between DGLAP and BFKL equations has been
found in  \protect\cite{Lipatov01,KoLi1,KoLi1b,KoLi2} for the first two orders of 
perturbation theory. These results can be useful for possible future 
study of the corresponding approximate relations in QCD.}
\par
The simple models in eqs. (\ref{fixsumma}) and (\ref{Gslut}), for
fixed and running couplings respectively, interpolate smoothly between
large and small $k_\perp$-values. They contain the essential features
of the CCFM and LDC models, and can therefore give an intuitive
picture of the transition between these two regimes. The transition
occurs for a fixed ratio between $\log k_\perp^2$ and $\log 1/x$ for
fixed coupling, and between $\log(\log k_\perp^2)$ and $\log 1/x$ for a
running coupling. A more extensive discussion, including effects of
non-leading terms in $\log 1/x$ and the properties of Laplace
transforms, is found in ref. \cite{LDCf}.
\subsection{The CCFM equation in the 
transverse coordinate representation}
\label{sec:ccfm-impact-parameter}
It might be useful to consider the CCFM equation using the transverse
 coordinate (or impact parameter)~\cite{Kwiecinski2002}
 $b$ conjugate to the transverse momentum 
$k_{\perp}$   
and  to introduce the $b$ dependent  gluon distribution 
$\bar {\cal A}(x,b,\Pmax^2)$ defined 
by the Fourier-Bessel transform of ${\cal A}(x,\kt^2,\Pmax^2)$: 
\begin{equation}
\bar {\cal A}(x,b,\Pmax^2) =  \int_0^{\infty}k_{\perp}dk_{\perp} 
J_0(bk_{\perp}){\cal 
A}(x,\kt^2,\Pmax^2). 
\label{bara}
\end{equation}
where $J_0(u)$ is the Bessel function.  The advantage 
of this representation becomes particularly evident in the so 
called 'single -loop' 
approximation that corresponds to the replacement 
$\bar q/z  \rightarrow \bar q$ and 
$\Delta_{NS} \rightarrow 1$.  In this approximation the CCFM equation reduces
 to the 
conventional DGLAP evolution and  eq.(\ref{CCFM_differential}) 
is diagonalized by the 
Fourier - Bessel transformation provided that we set the same argument 
$\mu^2 \sim \bar q^2$ 
in both terms in this equation.  The evolution equation for  
$\bar {\cal A}(x,b,\Pmax^2)$ 
 reads: 
\begin{equation}
\bar q^2 {d\over d\bar q^2 } {x\bar {\cal A}(x,b,\Pmax^2) \over \bar 
\Delta_s(\Pmax^2,Q_0^2)}=
\bar \alpha_S(\bar q^2)\int dz \bar P(z) J_0[(1-z)\bar q b] 
{x^{\prime} \bar {\cal A}(x^{\prime},b,\Pmax^2)\over 
\bar \Delta_s(\Pmax^2,Q_0^2)}
\label{evb1}
\end{equation} 
where 
\begin{equation}
\bar P(z)={1\over 1-z} -2 +z(1-z)+{1\over z}
\label{pz}
\end{equation}
In equation (\ref{evb1}) the argument $\mu^2$ of the QCD coupling was set
 $\mu^2 =\bar q^2$ 
and 
\begin{equation}
 \bar \Delta_s(\Pmax^2,Q_0^2)=\exp
 \left(-\int_{Q_0^2}^{\bar q^2} {dq^2\over q^2} 
\int_0^{1-Q_0/q}dz \bar \alpha_S(q^2) z \bar P(z)\right)
\label{sudbar} 
\end{equation}
In the splitting function $\bar P(z)$ we have included the terms which are 
non-singular at $z=0$ or $z=1$ in order to get a complete DGLAP evolution
equation for  
the integrated gluon distribution. For the same reason we have also 
extended the definition of the 
Sudakov form factor by including the complete splitting function $\bar P(z)$
 and not only its 
part which is singular at $z=1$ (cf. eq. (\ref{sudbar})). \\
Equation (\ref{evb1}) can be solved exactly after going to the moment
 space and 
introducing the moment function $\tilde {\cal A}_{\omega}(b,\Pmax^2)$ 
defined by: 
\begin{equation}
\tilde {\cal A}_{\omega}(b,\Pmax^2)= \int_0^1 dx x^{\omega} 
\bar {\cal A}(x,b,\Pmax^2) 
\label{moma}
\end{equation}
The solution of the evolution equation for the moment function 
$\tilde {\cal A}_{\omega}(b,\Pmax^2)$ reads: 
\begin{equation}
\tilde {\cal A}_{\omega}(b,\Pmax^2)=\tilde {\cal A}_{\omega}^0(b) \cdot
\exp \left[\int_{Q_0^2}^{\bar q^2}{dq^2\over q^2} \bar 
\alpha_S(q^2)\int_0^{1-Q_0/\bar q}
dz z\bar P(z)(z^{\omega-1} J_0(bq(1-z))-1)\right] 
\label{solmoma}
\end{equation}
where  $\tilde {\cal A}_{\omega}^0(b)$ is the Fourier-Bessel transform
 of the (input)  
unintegrated distribution at $\bar q=Q_0$. 
 It should be noted that at $b=0$,   
$\tilde {\cal A}_{\omega}(b,\Pmax^2)$  is proportional to the moment
 function of the integrated gluon 
distribution $g(x,\bar q)$. The solution (\ref{solmoma}) at $b=0$ reduces 
to the 
solution of the LO DGLAP equation for the moment function of the integrated 
gluon 
distribution.  It should also be noted that the integrand in  
eq.(\ref{solmoma})  is free from singularities at $z=1$ and so we can set 
$Q_0=0$ in the upper 
integration limit over $dz$. The formalism presented above is similar to 
that used for the description of the 
transverse momentum  distributions in (for instance) the Drell-Yan process 
(see e.g. \cite{\PT}).  
\\
In order to obtain more insight into the structure of the unintegrated 
distribution which follows from the CCFM equation in the single loop 
approximation 
it is useful to adopt in eq.(\ref{bara}) the following approximation of
 the Bessel function:
\begin{equation}
J_0(u) \simeq \Theta(1-u)
\label{apprj0}
\end{equation}
which gives: 
\begin{equation}
 {\cal A}(x,k^2_{\perp},\Pmax^2) \simeq 
 2{d \bar {\cal A}(x,b=1/k_{\perp},\Pmax^2)\over dk_{\perp}^2} 
\label{deriv}
\end{equation}
Using the same approximation for the Bessel function in eq.(\ref{solmoma})
we get, after some rearrangements, the following 
relation between unintegrated and integrated gluon distributions: 
\begin{equation}
{\cal A}(x,\kt^2,\Pmax^2) \simeq 
{d\over dk_{\perp}^2}[g(x,k_{\perp}^2) \bar \Delta_s(\Pmax^2, \kt^2)]
\label{apsol}
\end{equation}
Similar relation has also been used in refs.~\cite{DDT,Kimber:1999xc} 
(see also eq.(\ref{DGLAP_unintegrated}) in the next section).\\
 In the general 'all loop' case, the non-Sudakov form-factor 
generates contributions which are no longer diagonal in the $b$ space 
and so the merit of using this representation is less apparent.      
However  in the leading $\log(1/x)$ approximation at small $x$  
the  CCFM equation  reduces  to  the BFKL equation 
with no scale dependence and 
the kernel of the BFKL 
equation in  $b$ space is the same as that in the (transverse) momentum space. 
The work to explore the CCFM equation in  $b$ space beyond the 
'single loop' and BFKL approximations is in progress.

\subsection{Available parameterizations of unintegrated gluon distributions}

Depending on the approximations used, different unintegrated gluon
distributions can be obtained (see eq.(\ref{intglu})). Here we describe
some of the so far published ones and make some
comparisons.\footnote{Parameterizations of the unintegrated gluon
  distributions described here are available as a FORTRAN code from
  {\tt http://www.thep.lu.se/Smallx} } We use the following
  classification scheme: $x{\cal G}(x,\kt^2)$ describes DGLAP type unintegrated
  gluon distributions, $x{\cal F}(x,\kt^2)$ is used for  pure BFKL and 
  $x{\cal A}(x,\kt^2,\Pmax^2)$ stands for a CCFM type or any other type having two
  scales involved.
\begin{itemize}
\item {\bf Derivative of the integrated gluon density}\\
  Ignoring the fact that the unintegrated density may depend on two
  scales we can invert eq.(\ref{intglu}) which gives the unintegrated
  gluon density ${\cal G}(x,\kt^2)$:
   \begin{equation}
   x{\cal G}(x,\kt^2) = \left. \frac{d xg(x,\mu^2)}{d \mu^2}\right|_{\mu^2 =
   \kt^2}
   \end{equation}
   Here $xg(x,\mu^2)$ can be any of the parameterizations of the gluon density
   available.
\item {\bf GBW } (Golec-Biernat W\"usthoff ~\cite{GBW_1999})\\
  Within the color-dipole approach of~\cite{GBW_1999,GBW_1998}, 
  a simple parameterization of the unintegrated gluon density was obtained which
  successfully describes both inclusive and also diffractive scattering.
  The unintegrated gluon density is given by~\cite{GBW_1999}:
  \begin{equation}
  {\cal F}(x,\kt^2) = \frac{3 \sigma_0}{4 \pi^2 \as} R_0^2(x) \kt^2
  \exp \left( -R_0^2(x)\kt^2 \right)
  \end{equation}
  with $ \sigma_0 = 29.12$~mb,  $\as = 0.2$ 
  and 
  $$R_0(x)=\frac{1}{Q_0}\left(\frac{x}{x_0}\right)^{\lambda/2}$$
  and $Q_0=1$~GeV, $\lambda=0.277$ and $x_0 = 0.41 \cdot 10^{-4}$, where the
  parameters correspond to the 
  parameterization which includes charm~\cite{GBW_1998}. 
\item {\bf RS } (Ryskin, Shabelski~\cite{ryskin_shabelski})\\
  A parameterization of the unintegrated gluon density satisfying the BFKL
  equation but including also the non-singular parts of the gluon splitting
  function as given by DGLAP  
  is presented in \cite{ryskin_shabelski}. 
  The integrated gluon density $xg(x,\mu^2)$ 
  is defined by:
  \begin{equation}
  xg(x,\mu^2) = \frac{1}{4 \sqrt{2}\pi^3} \int_0^{\mu^2} \phi(x,\kt^2) d\kt^2
  \end{equation}
  The function $\phi(x,\kt^2)$ is obtained by a solution of the 
  evolution equation (including LO BFKL and the remaining part of the 
  DGLAP splitting function, but
  without applying the {\it kinematic constraint})
  in the perturbative region $\kt^2 > Q_0^2$. The non-perturbative part of the
  gluon density is identified as the collinear gluon density 
  $xg(x,Q_0^2)$ at a small scale
  $Q_0^2 \sim 4 $~GeV$^2$:
  \begin{equation}
  {\cal F}(x,\kt^2) = \left\{\begin{array}{ll}
     \frac{xg(x,Q_0^2)}{Q_0^2} & \mbox{if  } \kt^2 < Q_0^2 \\
     \frac{\phi(x,\kt^2)}{4 \sqrt{2}\pi^3} & \mbox{if  } \kt^2 \geq Q_0^2  
  \end{array} \right.
  \end{equation}
  An explicit form of the parameterization of the function $\phi$ 
  together with the fitted parameters
  are given in~\cite{ryskin_shabelski}.
 \item {\bf KMS } (Kwiecinski, Martin, Stasto~\cite{Martin_Stasto})\\
   In the approach of {\it KMS}~\cite{Martin_Stasto} 
   three modifications to the BFKL
   equation are introduced in order to extend its validity to cover
   the full range in $x$. Firstly, a term describing the leading order
   DGLAP evolution is added. The inclusion of this term has the effect
   of softening the small $x$ rise and to change the overall
   normalization.  Secondly, the kinematic constraint eq.(\ref{eq:CC}) 
   is applied, which has its origin in
   the requirement that the virtuality of the
   exchanged gluon is dominated by its transverse momentum.
   Thirdly, the BFKL equation is solved only in the region of $\kt > \kti{0}$,
   whereas the non-perturbative region is provided by the 
   collinear gluon density $xg(x,\kti{0}^2)$ at the
   scale $\kti{0}$. 
   The border between the perturbative and
   non-perturbative regions has been taken to be $\kti{0}=1$~GeV$^2$.
   \par
   Also a term which allows the quarks to contribute to the evolution
   of the gluon has been introduced. The single quark contribution is
   controlled through the $g \to q\bar{q}$ splitting. 
   The input gluon and quark distributions to this {\it unified} DGLAP-BFKL
   evolution equation are determined by a fit to HERA $F_2$ data. The
   unintegrated gluon density ${\cal F}(x,\kt^2)$
   is still only a function of $x$ and $\kt^2$, which strictly satisfies:
   \begin{equation}
   xg(x,Q^2) = \int^{Q^2} \frac{d\kt^2}{\kt^2} f(x,\kt^2) = 
   \int^{Q^2} d\kt^2 {\cal F}(x,\kt^2).
   \end{equation}
 \item {\bf JB } (J.~Bl\"umlein~\cite{Bluemlein}) \\
   In the approach of
   {\it JB}~\cite{Bluemlein} the general $\kt$-factorization formula:
   $$ \sigma(x,\mu^2)=\int d\kt^2 \hat{\sigma}(x,\kt^2,\mu^2)\otimes {\cal
   A}(x,\kt^2,\mu^2)$$
   is rewritten in the form~\cite{CE}:
   \begin{equation}
   \sigma(x,\mu^2)= \sigma^0(x,\mu^2)\otimes g(x,\mu^2) + 
   \int_0^{\infty} d\kt^2 \left(\hat{\sigma}(x,\kt^2,\mu^2)
    -\sigma^0(x,\mu^2) \right)
   \otimes {\cal A}(x,\kt^2,\mu^2),
   \label{CE_fact}
   \end{equation} 
   which gives the relation:
   $$  g(x,\mu^2) = \int_0^{\infty} d\kt^2 {\cal A}(x,\kt^2,\mu^2).$$
   In the case of eq.(\ref{CE_fact}) the \kt\ dependent gluon
   distribution satisfying
   the BFKL equation can be represented as the convolution of the
   integrated gluon density $xg(x,\mu^2)$ (for example GRV~\cite{GRV95})
   and a universal function ${\cal B}(x,\kt^2,\mu^2)$:
   \begin{equation} \label{conv}
     {\cal A} (x,\kt^2,\mu^2) = \int_x^1 
     {\cal B}(z,\kt^2,\mu^2)\, 
     \frac{x}{z}\,g(\frac{x}{z},\mu^2)\,dz, 
   \end{equation} 
  The universal function ${\cal B}(x,\kt^2,\mu^2)$ can be represented as a series
(see~\cite{Bluemlein}):
$$\sum_{i=1}^{N=\infty} d_i \tilde I_{i-1}(...),$$
where $\tilde I_i=I_i$ if $\kt^2 >\mu^2$ and 
$\tilde I_i=J_i$ if $\kt^2 <\mu^2$,
respectively and $J_i$ and $I_i$ are  Bessel functions
for real and imaginary arguments.
The series comes from the expansion of the BFKL anomalous dimension
with respect to $\as$. 
The first term of the above expansion explicitely describe BFKL dynamics in 
the  double-logarithmic approximation:
   \begin{equation} \label{J0} 
     {\cal B}(z,\kt^2,\mu^2)=\frac{\alb}{z\,\kt^2}\, 
     J_0(2\sqrt{\alb\log(1/z)\log(\mu^2/\kt^2)}), 
     \qquad \kt^2<\mu^2, 
   \end{equation} 
   \begin{equation}\label{I0} 
     {\cal B}(z,\kt^2,\mu^2)=\frac{\alb}{z\,\kt^2}\, 
     I_0(2\sqrt{\alb\log(1/z)\log(\kt^2/\mu^2)}), 
     \qquad \kt^2>\mu^2, 
   \end{equation} 
   where $J_0$ and $I_0$ are the standard Bessel functions
   (for real and imaginary arguments, respectively) and 
   $\alb=3 \as/\pi$ is connected to the pomeron intercept $\Delta$
   of the BFKL equation in LO $\omega_{\mathrm{LL}}=\alb 4 \log 2$. 
   An expression for 
   $\omega_{\mathrm{NLL}}$ in NLO is given in~\cite{\BFKLNLO}: 
   $\omega_{\mathrm{NLL}}=\alb 4 \log 2- N\alb^2$.
   Since the equations behave infrared finite no
   singularities will appear in the collinear approximation for small
   $\kt$. 
   
\item {\bf KMR } (Kimber, Martin, Ryskin~\cite{martin_kimber})\\
  In {\it KMR}~\cite{martin_kimber} the dependence of the unintegrated gluon
  distribution on the two scales $\kt$ and $\mu$ was investigated: the scale
  $\mu$ plays a dual role, it acts as the factorization scale and also controls
  the angular ordering of the partons emitted in the evolution. Already in the
  case of DGLAP evolution, unintegrated parton distributions as a function of
  the two scales were obtained by investigating separately the real and virtual
  contributions to the evolution. In DGLAP the unintegrated gluon density ${\cal
  A}(x,\kt^2,\mu^2)$ can be written as:
  \begin{eqnarray}
  x{\cal A}(x,\kt^2,\mu^2) & = & T(\kt^2,\mu^2)\frac{1}{\kt^2}
   \frac{\as(\kt)}{2 \pi } 
             \int_x ^{(1-\delta)} P(z) x'g(x',\kt) dz
                                  \label{DGLAP_unintegrated}
\\
                 & = & 
                  T(\kt^2,\mu^2) \left. \frac{d xg(x,\mu^2)}{d \mu^2} 
			\right|_{\mu^2=\kt^2} \nonumber
  \end{eqnarray}
  where $T(\kt^2,\mu^2)$ is the Sudakov form factor, resumming the virtual
  corrections:
  $$
  T(\kt^2,\mu^2)=\exp{\left(-\int_{\kt^2}^{\mu^2} \frac{\as(\kt)}{2 \pi} 
  \frac{d \kt^{'2}}{d \kt^{'2}} \int_0^{(1-\delta)} P(z')  
  dz'
  \right)}$$
  The structure of eq.(\ref{DGLAP_unintegrated}) is similar to the 
  differential form of the CCFM
  equation in eq.(\ref{CCFM_differential}), but with essential differences in
  the Sudakov form factor as well as in the scale argument in 
  eq.(\ref{DGLAP_unintegrated}).
  The splitting function $P(z)$ in eq.(\ref{DGLAP_unintegrated}) is now
  taken from the single scale evolution of the {\it unified} DGLAP-BFKL
  expression discussed before in {\it KMS}~\cite{Martin_Stasto}.
  As in {\it KMS} the unintegrated gluon density
  $f(x,\kt^2,\mu^2)$ starts only for 
  $\kt^2 > \kti{0}^2=1$~GeV$^2$. 
  An extrapolation to cover the whole range in $\kt^2$ has
  been performed~\cite{Kimber_pc_2001} such, that:
  \begin{equation}
  x{\cal A}(x,\kt^2,\mu^2) = \left\{\begin{array}{ll}
        \frac{xg(x,\kti{0}^2)}{\kti{0}^2} & \mbox{if  }\kt < \kti{0} \\
     \frac{f(x,\kt^2,\mu^2)}{\kt^2} & \mbox{if  } \kt \geq \kti{0}   
  \end{array} \right.
  \end{equation}
with $xg(x,\kti{0}^2)$ being the integrated MRST~\cite{MRST}
gluon density function. The
unintegrated gluon density $x{\cal A}(x,\kt^2,\mu^2)$ therefore is
normalized to the MRST function when integrated up to $\kti{0}^2$. 

\item {\bf JS} (Jung, Salam~\cite{jung_salam_2000,CASCADE})\\
  The CCFM evolution equations have been solved 
  numerically~\cite{jung_salam_2000,CASCADE} using a Monte Carlo method. 
  The complete partonic evolution
  was generated according to eqs.(\ref{CCFMa}-\ref{ns_new}) 
  folded with the off-shell matrix elements for boson gluon fusion.
  According to the
  CCFM evolution equation, the emission of partons during the initial
  cascade is only allowed in an angular-ordered region of phase space.
  The maximum allowed angle $\Xi$ for any gluon emission sets the
  scale $\Pmax$ for the evolution and is defined by the hard
  scattering quark box, which connects the exchanged gluon to the
  virtual photon (see section~\ref{sec:CCFM_evolution}). 
  \par
  The free parameters of the starting gluon distribution were fitted to
  the structure function $F_2(x,Q^2)$ in the range $x < 10^{-2}$ and 
  $Q^2 > 5$~GeV$^2$ as described
  in~\cite{jung_salam_2000}, resulting in the CCFM unintegrated gluon
  distribution $ x{\cal A}(x,\kt^2,\Pmax^2) $. 
\end{itemize}
\par
\begin{figure}[htb]
  \begin{center} 
    \vspace*{1mm} 
    \epsfig{figure=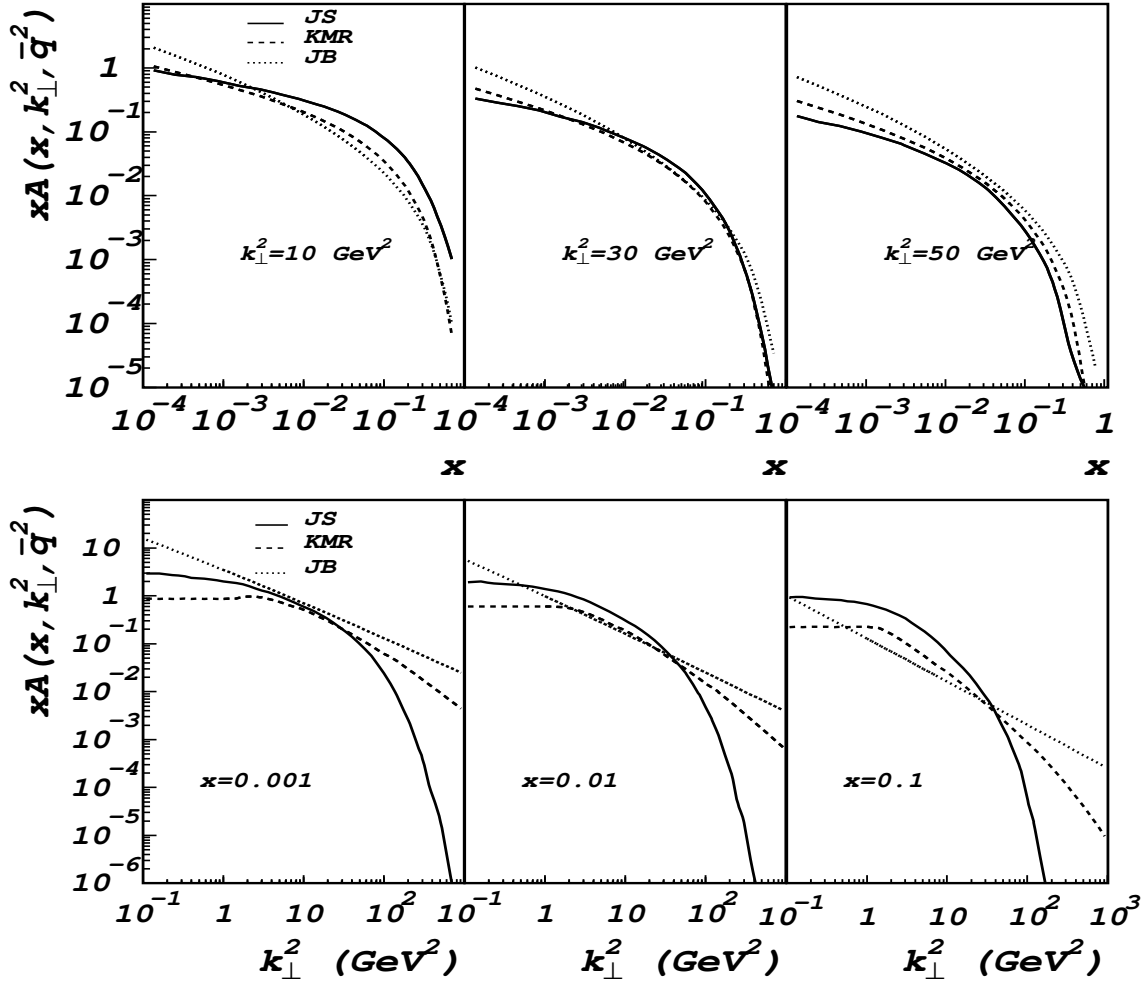, 
      width=17.5cm,height=15cm} 
    \caption{{\it  
        The unintegrated, two scale dependent gluon distributions at 
          $\Pmax=\mu=10$~GeV 
        as a function of $x$ for different values of $\kt^2$
        (upper) and as a function of $\kt^2$ for different values of
        $x$ (lower): JS~\protect\cite{CASCADE,jung_salam_2000} 
          (solid line), KMR~\protect\cite{martin_kimber} (dashed 
        line) and JB~\protect\cite{Bluemlein} (dotted
        line)}\label{gluon_twoscale}}
  \end{center}
\end{figure}

In the following we show comparisons between the different
unintegrated gluon distributions.  The {\it JS} unintegrated gluon
density serves as our benchmark, because calculations based on this
have shown good agreement with experimental measurements, both a HERA
and at the TEVATRON.  In Fig.~\ref{gluon_twoscale} we show the gluon
distributions at $\mu=\Pmax  = 10$ GeV obtained from the different BFKL
sets ({\it KMR}, {\it JB}) as a function of $x$ for different
values of $\kt^2$ and as a function of $\kt^2$ for different values of
$x$ together with the {\it JS} unintegrated gluon density. It is
interesting to note, that the {\it JS} gluon is less steep at small $x$
compared to the others. However, after convolution of the {\it JS} gluon with
the off-shell matrix element, the scaling violations of $F_2(x,Q^2)$
and the rise of $F_2$ towards small $x$ is reproduced, as shown in
\cite[Fig.~4 therein]{jung_salam_2000}. In the lower part of
Fig.~\ref{gluon_twoscale}, the effect of the evolution scale on the
distribution in $\kt^2$ is shown: The {\it JS} and {\it KMR} sets
both consider angular ordering for the evolution, whereas in {\it
  JB} the evolution in $\kt$ is decoupled from the evolution in
$\mu$, resulting in a larger $\kt$ tail. {\it JS} also includes
exact energy-momentum conservation in each splitting which further
suppresses the high-\kt\ tail.  Fig.~\ref{gluon_twoscale} also shows,
that the small $\kt^2$ region, which essentially drives the total
cross sections, is very different in different parameterizations.
\par
\begin{figure}[htb] 
  \begin{center} 
    \vspace*{1mm} 
    \epsfig{figure=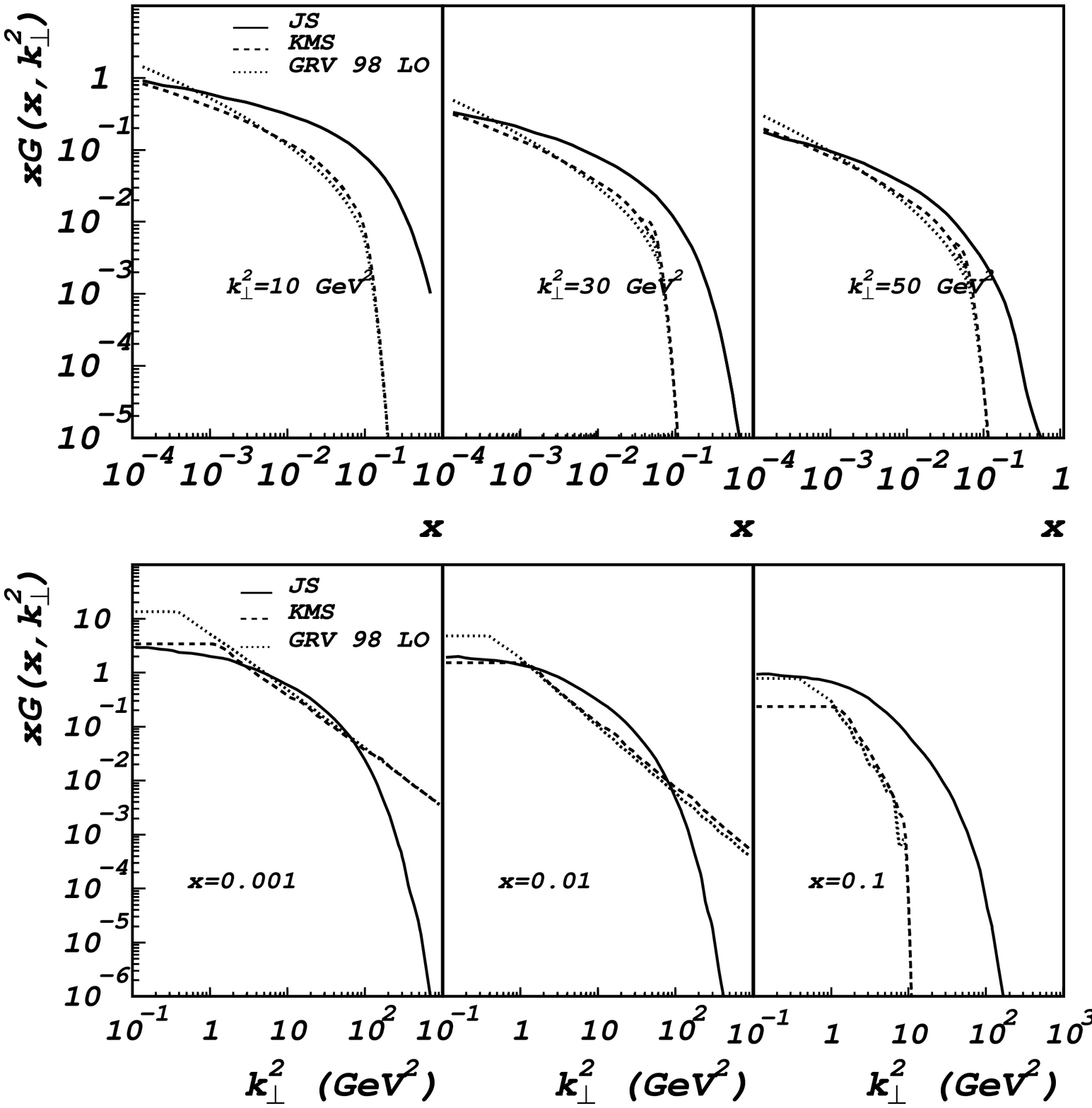,
      width=17.5cm,height=15cm}
    \caption{{\it
       The unintegrated, one scale gluon distributions   
        as a function of $x$ for different values of $\kt^2$
        (upper) and as a function of $\kt^2$ for different values of
        $x$ (lower):  KMS~\protect\cite{Martin_Stasto} (dashed line)
          and  derivative GRV $LO$ (dotted line)
	  here compared to the {\it two} scale gluon distribution of 
          JS~\protect\cite{CASCADE,jung_salam_2000} (solid line)
	  at $\Pmax=\mu=10$~GeV  (same as in Fig.\protect\ref{gluon_twoscale})
        }\label{gluon_onescale}}
  \end{center}
\end{figure}
\begin{figure}[htb] 
  \begin{center} 
    \vspace*{1mm} 
    \epsfig{figure=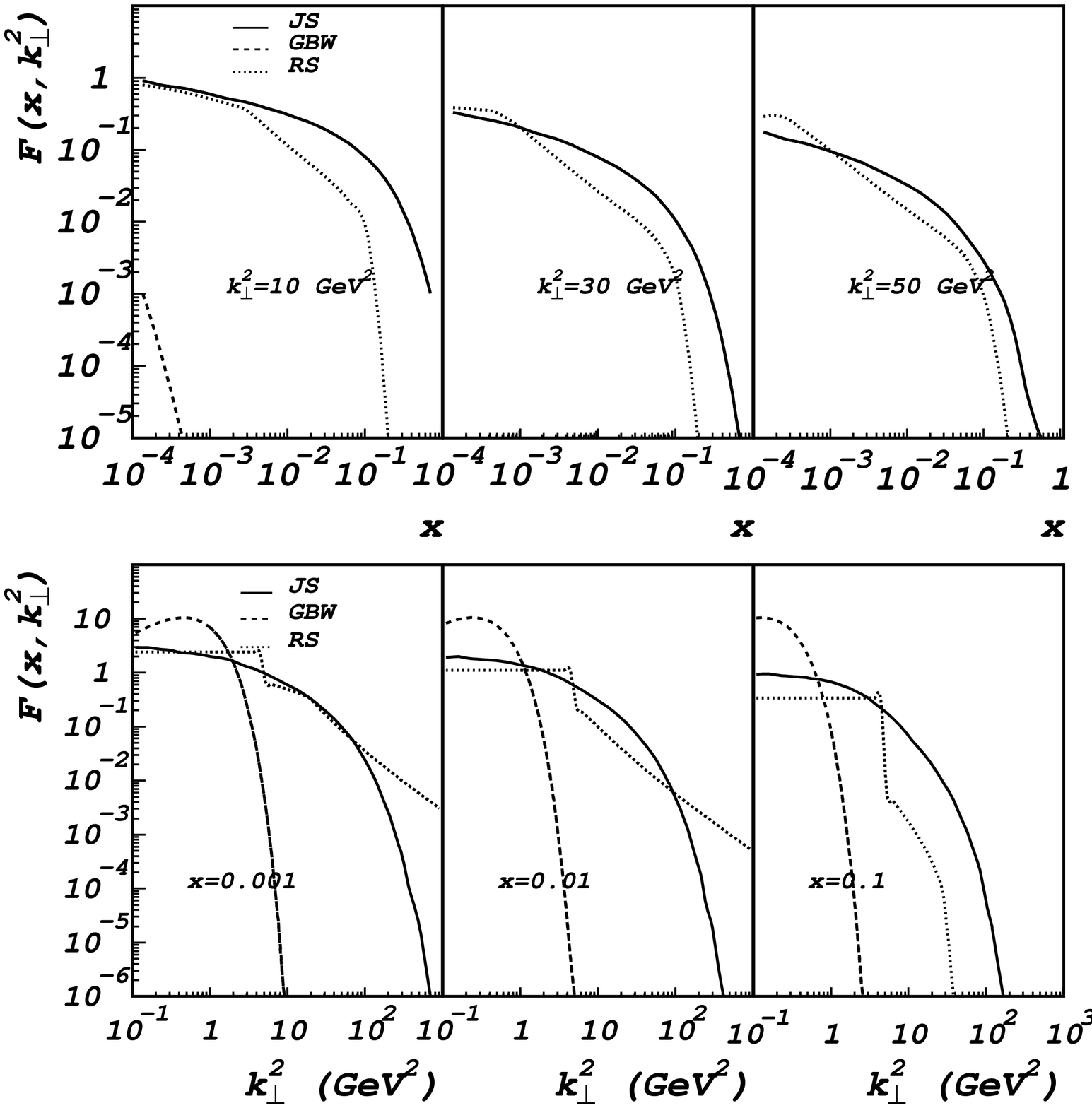,
      width=17.5cm,height=15cm}
    \caption{{\it
       The unintegrated, one scale gluon distributions   
        as a function of $x$ for different values of $\kt^2$
        (upper) and as a function of $\kt^2$ for different values of
        $x$ (lower): GBW~\protect\cite{GBW_1999} (dashed line)
          and  RS~\protect\cite{ryskin_shabelski} (dotted line)
	  here compared to the {\it two} scale gluon distribution of 
          JS~\protect\cite{CASCADE,jung_salam_2000} (solid line)
	  at $\Pmax=\mu=10$~GeV (same as in Fig.\protect\ref{gluon_twoscale})
        }\label{gluon_onescale2}}
  \end{center}
\end{figure}

In Fig.~\ref{gluon_onescale} we show a comparison of the gluon density
distribution obtained from the derivative method (using GRV98 LO) 
and {\it KMS} 
and compare it to the {\it JS} gluon at $\Pmax=10$ GeV. 
The {\it KMS} and ``derivative of GRV"
give very similar unintegrated gluon distributions, which is a result
of the strict relation between the collinear (integrated) and unintegrated
gluon distributions, that was used in {\it KMS}. One also has to note, that 
{\it KMS} and GRV98 use very similar data sets from HERA for their fits. 
In Fig.~\ref{gluon_onescale2} we show a comparison of the 
unintegrated gluon density parameterizations from {\it GBW} and {\it RS} 
and compare it to the {\it JS} gluon at $\Pmax=10$ GeV. The {\it RS} set is shown
only for historical reasons, since it was one of the first unintegrated gluon
distributions available. The {\it GBW} unintegrated gluon density, 
although successful in describing
inclusive and diffractive total cross sections at HERA, is suppressed
for large $\kt$ values (see Fig.~\ref{gluon_onescale2}),
which is understandable, since large $\kt$ values 
can only originate from parton evolution, but were not treated in {\it GBW}.
However, it is  interesting to compare it also to
more exclusive measurements. One can also see, that the  {\it GBW} gluon
decreases for $\kt \to 0$.
\begin{table}[htb]
\begin{center}
\begin{tabular}{|l|c|c|c|c|}\hline
        & \multicolumn{4}{|c|} {$\sigma [\mbox{nb}]$ } \\ \hline
        & {\it JS} & {\it KMR} & {\it JB} & {\it GBW} \\ \hline
$e p \to e' c \bar{c} X$ ($Q^2 < 1$ GeV$^2$) & 696.8 & 412.8 & 741.3 & 735.3 \\ \hline
$e p \to e' c \bar{c} X$ ($Q^2 > 1$ GeV$^2$) & 80.2 & 47.4 & 87.7  & 82.0 \\ \hline
$e p \to e' b \bar{b} X$ ($Q^2 < 1$ GeV$^2$) & 5.36 & 2.78 & 4.38 & 5.64 \\ \hline
$e p \to e'  X$ ($Q^2 > 1$ GeV$^2$)          & 838.7 & 610.0  & 6550.0  & 4134.0  \\ \hline
$e p \to e'  X$ ($Q^2 > 5$ GeV$^2$)          & 212.8 & 127.2  & 585.1  &  564.0 \\ \hline
$p \bar{p} \to  b \bar{b} X$ ($\sqrt{s}=1800$ GeV) 
&88100. & 27489. & 78934.  & 65990. \\ \hline
\end{tabular}
\caption{\it Cross sections for different processes at HERA and TEVATRON using
different parameterizations of the unintegrated gluon distributions. The cross
sections are calculated with the \CASCADE ~\protect\cite{CASCADEMC} Monte Carlo
generator. In all cases the one-loop $\alpha_s(\mu^2)$ is used with 
$\mu^2=p_{\perp}^2 + m_q^2$, where $p_{\perp}$ is the transverse momentum of one of the quarks
in the partonic center-of-mass frame and $m_q=0.140, 1.5, 4.75$~GeV is 
mass of the
light, charm and bottom quarks.
}\label{xsections}
\end{center}
\end{table} 
\par
In Table~\ref{xsections} we present cross sections calculated with
four of the different
unintegrated gluon distributions discussed above for 
 heavy flavor production and inclusive deep inelastic
scattering at HERA energies
($\sqrt{s}=300$ GeV)
as well as the total bottom cross section at the TEVATRON. The
benchmark is again the {\it JS} parameterization, which is able to describe
the corresponding measurements at HERA and at the TEVATRON reasonably well.
Large variations in the 
predicted cross sections are observed. 
It is clear that the parameterizations of the unintegrated gluon
distributions are very poorly constrained, both theoretically and
experimentally. The differences in definition makes it difficult to go
beyond a purely qualitative comparison between the different
parameterizations. Also, since even the integrated gluon density is
only indirectly constrained by a fit to $F_2(x,Q^2)$, it may be necessary to
look at less inclusive quantities to get a good handle on the
unintegrated gluon.
\subsection{Beyond leading logarithms}
\label{subsec:beyond-leading-order}
In this section the attempts to go beyond leading order are described and
summarized. First, general aspects of next-to-leading effects are discussed 
and then the CCFM approach is critically considered,
\par
We write the momenta  of the $t$-channel gluons 
(in Sudakov representation) as 
$k_i = x_{i}^+ P_1 + x_{i}^- P_2 + \kti{i}$  and the emitted gluons as 
$p_i = v_{i} (P_1 + \xi_i P_2) + p_{\perp i }$ (eq.(\ref{ccfm_q})). 
The discussion in this section will be based on the strong ordering limit,
that is 
$x^+$, angles and $x^-$ are strongly ordered, which means that factors of
$(1-z)$ can be safely ignored since they are of ${\cal O}(1)$ for $z \to 0$.
In the high energy limit with strong ordering, one can talk of  $x^+$ ordering 
of exchanged gluons  or of  
angular ordering of the emitted gluons. The 
variable $\xi$ (related to the angle of the emitted gluons)  can be
written (in this approximation) as: 
\begin{equation}
\sqrt{\xi_i}  = \frac{p_{\perp i}}{x_{i-1}^+ \sqrt{s}}
=\sqrt{\frac{x_{i-1}^+}{x_{i-1}^-}},
\end{equation}   
with $v_i \sim x_{i-1}^+$ for $z \to 0$ (eq.(\ref{ccfm_q})). 
The last expression in the above
equation is obtained from:
\begin{equation}
x_{i-1}^- \sim v_i \xi_i \sim x_{i-1}^+
\left(\frac{p_{\perp i}}{x_{i-1}^+ \sqrt{s}}\right)^2.
\label{xplusxminus}
\end{equation}

\subsubsection{General next-to-leading order investigations}
\label{susubsec:NLLinvestigations}
It is known from many 
contexts of QCD that for reasonably accurate predictions it is
necessary (at the very least) to go to next-to-leading order. In what
follows, one has to remember that the role of next-to-leading
corrections in the
$\kt$-factorization approach is very different to the ones in the collinear
approach, since part of the standard NL corrections are already included at LO
level in $\kt$-factorization, as was discussed in 
chapter~\ref{sec:shell-matr-elem}.
\par
For a next-to-leading logarithmic (NLL) calculation of a cross section
at high energies, there are two main ingredients. One is the NLL
corrections to the
`kernel' of the BFKL equation, generating terms $\as (\as \log 1/x)^n$.
This part should be independent of the process under consideration.
The second part is the correction to the impact factors (off-shell
matrix elements) at either end, and is the source of the process
dependence in the NLL corrections. Processes of interest include
$\gamma^*\gamma^*$ scattering in electron positron annihilation,
forward jets at HERA and Mueller-Navelet jets at hadron-hadron colliders.
In order to describe these processes at NLO level of accuracy,
one needs the photon impact factor and the jet vertex. For both elements
NLO calculations are on the way: for the photon impact factor the
main are the off-shell matrix elements described in 
section~\ref{susubsec:NLLme} 
(refs. \cite{bartels-nlo-qqbar-virtual,bartels-nlo-qqbar-real_1,%
bartels-nlo-qqbar-real_2}), and the quark induced jet vertex has recently been
computed in \cite{BCV2001}.
\par
One of the major developments in past years has been the completion of
the calculation of the NLL corrections to the BFKL kernel. This was a
significant enterprise, lasting almost a decade
\cite{\BFKLNLO}. Once the various contributions have been assembled,
the final result can be given in a fairly compact form
\cite{\BFKLNLO}.  It can be summarized through the following relation
between the next-to-leading BFKL power, $\omega_{\mathrm{NLL}}$, 
 and the leading power $\omega_{\mathrm{LL}}=\alb 4 \log 2$:
\begin{equation}
  \label{eq:NLLsummary}
  \omega_{\mathrm{NLL}}=\alb 4 \log 2- N\alb^2  
  \simeq \omega_{\mathrm{LL}} (1 - 6.2 \as)\,.
\end{equation}
Substituting a typical value for $\as$, say $0.2$, one finds a
negative power --- so rather than improving the accuracy of the
predictions, the NLL corrections seem to lead to nonsensical answers.
A more detailed analysis suggests that for problems involving two
substantially different transverse scales, the inclusion of the NLL
corrections leads to an even worse problem, namely negative cross
sections \cite{Ross98}. So at first sight it seems that the perturbation
series for BFKL physics is simply too poorly convergent for it to be
of any practical use.
\par
Despite this, there are indications that ways exist of using
the NLL corrections for phenomenological purposes
($\gamma^*\gamma^*$, forward jets or TEVATRON jets with large rapidity
separation
might be examples for this).
This is because it
is possible to identify a well-defined physical origin for large parts
of the NLL terms. These parts can then be calculated at all orders,
and the remaining pieces are then a much smaller NLL correction.
\par
A clue as to the origin of the large corrections can be obtained by
examining their structure in the collinear limit (where one transverse
scale is much larger than the other). 
From DGLAP we think we understand the origin of all terms involving 
$(\alpha_s \log Q^2)^n$ -- they are associated with strong orderings 
in $\kt$. So e.g. in the BFKL NLO corrections (terms 
$\propto \alpha_s^2$) we have a piece with 
$ (\alpha_s^2 \log^2 Q^2)$ which is something we already know
about from DGLAP. Terms with this ``collinear'' enhancement
(a number of powers of log Q) turn
out to be responsible for over 90\% of the NLO corrections to
BFKL, and so are the reason for the large size of these
corrections. But since their origin is just DGLAP physics,
which we know well, we can also predict the terms that arise
at NNLO ($\as^3 \log^3 Q^2$) etc. and resum them. 
\par
Suppose one wishes to calculate
the high-energy behavior of a Green's function with a squared center
of mass energy $s$ and transverse scales $Q,Q_0$ at the two ends of
the chain.
It is convenient to examine this in Mellin transform space, with
$\omega$ conjugate to $s/(Q Q_0)$ and $\gamma$ conjugate to the squared
transverse momentum ratio $Q^2/Q_0^2$.
 One can then write the BFKL
kernel as
\begin{equation}
  \omega = 
  \asb \left(\chi_0 + \asb \chi_1 +
    \cdots \right)\,,\qquad \quad \asb=\frac{\as N_C}{\pi}\,.
\end{equation}
For small $\gamma$ (corresponding to a large ratio of transverse
momenta) the leading part of the BFKL kernel goes as
\begin{equation}
  \chi_0 \simeq \frac{1}{\gamma} + {\cal O}(\gamma^2)\,.
\end{equation}
In the same region the NLL corrections behave as
\begin{equation}
  \chi_1 \simeq 
  -\frac{1}{2\gamma^3} - \frac{11}{12\gamma^2}
    + {\cal O}(1)\,.
\end{equation}
The extra divergences at small $\gamma$ can be quite easily understood
because each power of $1/\gamma$ (after inverse Mellin transform) 
corresponds to a logarithm of transverse momentum ($\log Q$). 
So for example the term with $1/\gamma^2$, given
that it multiplies $\as^2$, corresponds to a contribution $(\as \log
Q^2/Q_0^2)^2$, and so looks like a standard term from DGLAP evolution.
One can verify the coefficient that would be expected from leading-log
DGLAP evolution and it comes out as exactly $-11/12$.
\par
The term proportional to $1/\gamma^3$ is slightly more subtle because
it looks super-leading compared to DGLAP. Recalling that it is
multiplied by $\asb^2$, we can see that it is related (after a Mellin
transform) to a term which we can write as %
$(\as \log Q^2/Q_0^2) \times (\as \log^2 Q^2/Q_0^2)$, i.e.\ a normal
DGLAP term, multiplied by a double log of $Q^2$. Such double
logs of $Q^2$ will be discussed in more detail in the next subsection.
Essentially, their presence or absence depends on whether
we resum high energy logs of $x^+$, $x^-$ or $\sqrt{x^+/x^-}$
 (or
equivalently $Q^2/s$, $Q_0^2/s$ or $Q Q_0/s$). The small-$\gamma$ part
of the NLO BFKL kernel has
been shown above with the high-energy logs defined in terms of
$\sqrt{x^+/x^-}$
 --- converting to the natural evolution variable for
DGLAP evolution, $x^+$, the $1/\gamma^3$ pole disappears, and one is
left with just the $1/\gamma^2$ term. In other words the small-$\gamma$
behavior is entirely constrained by known results from DGLAP. 
\par
It is important to note though that the $1/\gamma^3$ term is not purely
an artefact in that there is an analogous triple pole around $\ga=1$,
associated with transverse logarithms for $Q_0 \gg Q$, and there is no
convention for the definition of the high-energy logs which allows one
to get rid of the 
triple poles simultaneously at $\ga=0$ and $\ga=1$.
\par
Given that these  collinearly enhanced contributions have a
simple physical origin, namely the strong ordering in transverse momenta,
they can be quite
straightforwardly calculated not just to NLL, but to all orders. For
example the generalization of the $1/\ga^3$ and $1/(1-\ga)^3$ terms
can be obtained by implementing the kinematic constraint discussed
above. The generalization of the $1/\ga^2$ and $1/(1-\ga)^2$ terms
comes from the leading-log DGLAP kernel and from the running of the
coupling.
\par
Of course an understanding of the structure of the kernel around
$\ga=0$ does not formally tell us anything about $\ga=1/2$, which is
the region of interest for small-$x$ predictions. What is remarkable
though is that a collinear approximation, i.e.\ taking just the known
poles at $\ga=0$ and $\ga=1$,
\begin{equation}
  \chi_1^{\mathrm{coll}} = -\frac{1}{2\gamma^3} 
-\frac{1}{2(1-\gamma)^3} - \frac{11}{12\gamma^2} -
\frac{11}{6(1-\gamma)^2}\,, 
\end{equation}
approximates the full next-to-leading corrections to better than
$93\%$ over the whole range $0<\ga<1$. 
The above equation is just the expression in `$\gamma$'-space (the Mellin
transform of $Q$ space) of terms that we know from DGLAP.
\par
This suggests that a practical approach to `improving' NLL BFKL might
just be to include the higher-order collinear terms at all orders.
This has been done in \cite{Salam1,CC98b,CCS99}, and is found to
considerably improve the stability of the predictions for $\omega$,
giving a phenomenologically sensible value of $\omega$ in the range
$0.25$ to $0.3$ for
$\as\simeq 0.2$. 
These kinds of improvements are necessary
whenever exact NLO corrections are used,
because without them the result is very
unstable (large renormalization scale uncertainty, negative power
$\omega$ for $\alpha_s > 0.15$).
The same approach can be extended to the calculation
of Green's functions and physical cross sections, and work in this
direction is currently in progress \cite{CiafColfSalaStas}.
\par
There have been other approaches to supplementing the NLL corrections
with higher-order contributions, notably performing a BLM (Brodsky Lepage
Mackenzie) change of the scale for $\alpha_s$ \cite{BFKLP} and
enforcing rapidity separations  
between emissions \cite{Schmidt}.
The basic idea behind the BLM scale choice is that a large part of the NLO
corrections come about because in the LO calculation the
scale choice for $\alpha_s$ was unreasonable. 
This is reflected in
the NLO corrections by a large piece proportional to $\beta_0$, which
is one of the
parameters of the  $\beta$ function controlling the renormalization scale
dependence of $\alpha_s$. However, in a NLO calculation one obtains 
coefficients of $n_f$, and of $C_A$, but not of $\beta_0$ as such, 
so one can't identify what the actual coefficient of $\beta_0$ is. BLM
identifies the ``amount'' of $\beta_0$ with the entire $n_f$ part, effectively
guessing from the coefficient of $n_f$ the coefficient of $\beta_0$, 
and then saying that it should all be re-absorbed into a different 
choice of scale for $\alpha_s$ in the LO piece. 
When the authors of~\cite{BFKLP} 
tried this for NLO BFKL 
they found that it made the calculations marginally worse (even more unstable),
unless they chose a very specific renormalization scheme. They argue
that this scheme, which appears in certain other contexts involving
three-gluon couplings (though not particularly in a small-$x$ limit)
might be a more natural choice when applying the BLM procedure, and
find a value for $\omega$ of around $0.15$.
\par
The approach which enforces rapidity separations is motivated by the
observation that the approximations in the LL calculation for BFKL are
satisfied only when there is a large gap in rapidity between
successive emissions, but that one nevertheless integrates over all
possible separations between emissions (up to the limit where
emissions have the same rapidity), which is plainly in violation of
the initial approximations. One can choose instead to integrate up to
some (a priori arbitrary) minimum rapidity separation, $\Delta \eta$.
This modifies one's prediction at higher orders, and in particular can
partially mimic the NLL corrections. However its explicit merging with
the NLL corrections by Schmidt \cite{Schmidt} revealed a significant
remaining instability with respect to variations of $\Delta \eta$.
There is evidence though that when combined with the collinear
resummation described above, much of the instability with respect to
the variation of $\Delta \eta$ disappears \cite{ForshawRossSabioVera}.
Essentially both methods (collinear enhancement, rapidity
separation) cut out certain regions of phase space. With the
``collinear'' approach the piece cut out is well defined,
whereas with the rapidity cut, the amount cut out depends on
$\Delta \eta$. But when  the two
methods are put together, a large fraction of the phase space that
would be removed by a rapidity cut has already been removed
by the collinearly enhanced terms, so the effect of rapidity
cut is much reduced.
\subsubsection{Anomalous dimensions.}
So far the discussion has been most relevant to processes with two
similar hard scales. In the case of re-summed small-$x$ splitting
functions (anomalous dimensions) in addition to the issues discussed
above, there is a further subtlety related to \emph{iteration} of the
running of the coupling. It has been discussed with complementary
methods but similar conclusions in 
\cite{CCS99,Schmidt,CCS00},
\cite{Thorne99a,Thorne99b,Thorne01} and
\cite{Altarelli01}. Essentially the result is that 
for a small-$z$ splitting function at scale $Q^2$, the typical scale in the
evolution is considerably higher than $Q^2$
\footnote{The effective high scale could be responsible (see discussions
in \cite{Q2evo1,Q2evo2,Q2evo3}) for the good agreement
between the experimental data for the
$F_2$ structure function and the perturbative estimations in the 
small $x$ range.}. Similar results have been found 
before also in~\cite{DoShi,Rsmallx1,Rsmallx2,wong}  
in the framework of DGLAP, and in~\cite{BFKLP} in the framework of BFKL.
The splitting function is only independent of $Q^2$ at leading order. 
At NLO it is no longer scale
invariant because the NLO piece comes in with a different
relative weight according to $\as(Q^2)$, as do NNLO etc. When 
 deducing the small-$x$ enhanced part of the splitting
function then the NLO, NNLO, NNNLO (in the DGLAP hierarchy, $\as^{n+m}
\log^{n}Q^2 \Rightarrow N^m LO$) pieces which are small-$x$ enhanced 
need to be determined at all orders. And the
final `splitting function' then depends on $Q^2$ because 
the relative weights of all these pieces is different
according to the $Q^2$ value.
The inclusion of all these terms causes the $P_{gg}(z)$ splitting
function to grow as a power $1/z^{1+\omega_c}$ at small $z$. Initial
calculations of this dependence led to a large value for $\omega_c$
and consequent
incompatibility with the data (see e.g.\ \cite{Forte96}).
However these predictions are modified by two classes of large
effects: NLL contributions, similar to those discussed above, and also
corrections associated with the fact that the effective scale of
$\alpha_s$ relevant to small-$z$ splitting functions is substantially
larger than $Q^2$. After an involved calculation one finds that,
parametrically, this implies corrections to
$\omega_c$ of  order $\as^{5/3}, \as^{7/3},\ldots$ (which need to be
resummed) and leads to a
significant reduction in the small-$x$ power of the splitting function
compared to the fixed-coupling case, as well as increased stability
with respect to the NLL corrections. 
This means that the power $\omega_c$ for the small-$z$ dependence of
the splitting function, $1/z^{1+\omega_c}$, is smaller than one would
expect (i.e. it's not the same $\omega(\as(Q^2))$ that is 
calculated for a process such as $\gamma^*(Q) \gamma^*(Q)$) and
furthermore it sets in only at very small $z$ (below $10^{-5}$).
\par
An approach which also involves some of the elements discussed above
has been proposed and used in 
\cite{Altarelli:1999vw,Altarelli:2000mh} 
to study scaling violations in the HERA $F_2$ data~\cite{Adloff:2000qk}. 
In their approach the authors make use of a collinear
resummation around $\gamma=0$, however argue that this cannot also be
applied to $\gamma=1$, which implies that it is not possible to
predict the height of the minimum of the characteristic
function. Accordingly the asymptotic power $\lambda$ of the small-$x$
splitting function $\sim x^{-\lambda}$ must be fitted to the
data. Within this approach the authors find an improved agreement with
the $F_2$ data compared to a pure NLO DGLAP fit. The optimal value for
$\lambda$ depends on the details of how the resummation is
implemented, in one case being negative ($\lambda \simeq -0.25$),
while in the other it is of the order of $\lambda \simeq 0.2$ which
would also be consistent with other resummation approaches.

\subsubsection{CCFM and problems beyond leading logs}

The CCFM evolution involves three partonic variables, 
the light-cone momentum
fraction, the emission angle and the transverse momentum. Looking at the
case where a forward jet in DIS has a $\pt^2$ much smaller than $Q^2$,
the natural variable when going up in scale from the proton towards
the virtual photon is the positive light-cone momentum fraction,
 $x^+=Q^2/(s y) 
\sim x_{bj}$, giving rise to the standard double-logs,
$\left(\alb\log\frac{Q^2}{\pt^2}\log\frac{1}{x^+}\right)^n$, coming from
integrals of the form $\int_{\pt^2}^{Q^2}dq^2/q^2 \int_x^1 dz/z$.
Naively going from the photon side with the negative light-cone
momentum fraction (compare eq.(\ref{xplusxminus}))
$x^-=\pt^2/ (s y) = x^+\pt^2/Q^2$ will give us 
so-called {\it illegal} double-logs of $Q^2/\pt^2$
\begin{equation}
\left(\alb\log\frac{Q^2}{\pt^2}\log\frac{1}{x^-}\right)^n =
\left(\alb\log\frac{Q^2}{\pt^2}\log\frac{1}{x^+} +
\alb\log^2\frac{Q^2}{\pt^2}\right)^n\label{eq:illdoublelog}
\end{equation}
The wording 
{\it illegal} comes from the observation, that 
the renormalization group equations (DGLAP) tell us what
classes of logs we can expect (they are first order
differential equations so we expect only single logs of $Q^2$). When
we get a double log of $Q^2$ then this is illegal because it cannot
be compatible with the renormalization group. Subtleties, of course, arise
because different ways of writing the same expression can
lead to the double log being visible or not, and that is what
the above eq.(\ref{eq:illdoublelog}) describes.
CCFM uses the angle (or more specifically $\xi$) as evolution variable.
Using $\sqrt{\xi} = \frac{\pt }{x^+ \sqrt{s}}$, then  
the ordinary collinear evolution with
increasing $\pt$ and automatically decreasing $x^+$ results in angular ordering
or ordering in $\sqrt{\xi}$. If on the other hand the $\pt$'s 
decrease then a decrease in $x^+$ does not automatically imply an increase 
in angle -- so angular ordering becomes a stronger condition.
It is the angular ordering $\sqrt{\xi} \sim \sqrt{x^+/x^-}$ which is
responsible for the {\it illegal} double logs.
\par
In the original formulation of CCFM a kinematic constraint was noted:
$$ k_{i}^2  > z_{+i} k^2_{i-1} ,$$
which was derived already in  \protect\cite[eq.(2.11)]{BCM} and in
\protect\cite{PYTHIAPSb} in a frame where parton $i$ is along the
$z$ axis. If the transverse momenta \kti{i} are to be the dominant
contributions  
to the virtualities $k_{i}^2$ (as assumed in the derivation of the
CCFM equation) then the so-called 
{\it kinematical} or {\it consistency constraint} is
obtained:
\begin{equation}
  \label{eq:CC}
  \kti{i}^2>z_{+i}\kti{i-1}^2.
\end{equation}
To merely have a self-consistent evolution equation (no infrared
unsafety, proper 
real-virtual cancellations), which gives correct cross sections 
and final-state distributions at leading logarithmic accuracy, the 
consistency constraint is not needed in the CCFM equation.
The way the consistency constraint was originally written into the 
CCFM equation should not be used because it messes up real-virtual
cancellations as discussed below in 
section~\ref{subsec:beyond-leading-order-event-generators}.

\subsubsection{NLL corrections and event generators}
\label{subsec:beyond-leading-order-event-generators}
There are also aspects in which the NLL corrections may be able to
help guide the construction of event generators.
They give a hint about the kind of physical effects which are likely
to be important (splitting functions, kinematic constraint). There are
also situations when building an event generator in which one is faced
with two possible strategies, and sometimes the NLL corrections give
clues as to the direction to be followed.
\par
One example relates to the running of the coupling. A priori it is not
necessarily clear what scale to use and for simplicity, in various
phenomenological contexts, the scale has often been chosen to be
$k_{\prp i}^2$ (as related to the branching $(i-1)\to i$ in
figure~\ref{fig:variables}). 
However, in the full NLO calculation for BFKL a term appears
\begin{equation}
  -\asb^2 \beta_0 \ln \frac{p_{\prp i}^2}{\mu^2}\,.
\end{equation}
The interpretation is very simple: 
the right scale in the LO kernel is simply the 
emitted $p_{\perp}$ (because if included in the LO kernel
and then expanded in powers of $\alpha_s$, one obtains 
precisely such a term at NLO).
\par
But knowledge of the `ingredients' needed for an event generator is
not always sufficient. One generally has a basic equation for the
branching, such as the CCFM equation, but inserting the relevant
corrections such as to obtain the correct structure of logs at the end
is often not trivial, because the full set of logs comes out of the
\emph{iteration} of the branching. Furthermore some care is often
needed to maintain proper cancellation of real and virtual
corrections.
\par
For example in normal BFKL the correct cubic poles around $\ga=1$ can
be obtained by inserting the requirement $k_{i}^2 > z_i k_{i-1}^2$ for
each branching. This corresponds to symmetrizing evolution up in
transverse scale (with $x^+$ as the evolution variable) and evolution
down in transverse scale (with $x^-$).
\par
However implementing this constraint directly into CCFM leads to the
wrong coefficient for the $1/(1-\ga)^3$ term --- in other words the
symmetrization is not properly accomplished. One can partially solve
the problem by modifying the virtual corrections (non-Sudakov) form
factor
\begin{equation}
\ln \Delta_{ns}(z,q^2,k^2_{\perp}) = - \int
    \frac{d{q'}^2}{{q'}^2} \int^1_z \frac{dz'}{z'} \asb({q'}^2) 
    \Theta({q'}^2 - {z'}^2 {q}^2)
    \Theta(\kt^2 - {q'}^2) { \Theta\left(\frac{k^2}{q^2} - z'\right)}\,,
\end{equation}
where the last $\Theta$ function accounts for the kinematic constraint
in the virtual corrections. This will not make the cross section
exactly symmetric, but it
will ensure that for evolution downwards in scale the evolution
variable does at least correspond roughly to $x^-$. On the other hand
as far the final state structure is concerned, this `hack'~\cite{hacks} cuts out
certain regions of phase space that should be left in.
An a priori way of determining whether the `hacks' 
 are any good should
be to examine how they compare to exact NLO calculations (but 
that means determining their expansion to ${\cal O}(\as^2)$, which
isn't necessarily an easy task). 
\par
Another source of next-to-leading-log corrections is the gluon splitting function.
At very large energies, the $1/z$ term in $P_{gg}$, included in BFKL and CCFM,
will certainly be dominant. However, the question is whether the treatment of
just this term is sufficient at energies available at present colliders.
\par
The effect of the new small $x$ is best seen in forward jet production 
in deep inelastic scattering, where the contribution from 
typical DGLAP dynamics is suppressed. In such a process, using a Monte Carlo
simulation, the distribution of $z$-values can be studied, and the validity of
the small $x$ approximation can be checked.
\par
\begin{figure}[htb]
  \begin{center}
    \begin{minipage}{4cm}
      \begin{tabular}{l}
        $Q^2>10$ GeV$^2$ \\
        $E_{\Prp\;jet} > 5$ GeV \\
        $\eta_{jet} < 2.6$ \\
        $x_{jet}>0.036$\\ 
        $0.5<E^2_\Prp/Q^2<2$
      \end{tabular}
    \end{minipage}
    \begin{minipage}{8cm}
      \epsfig{figure=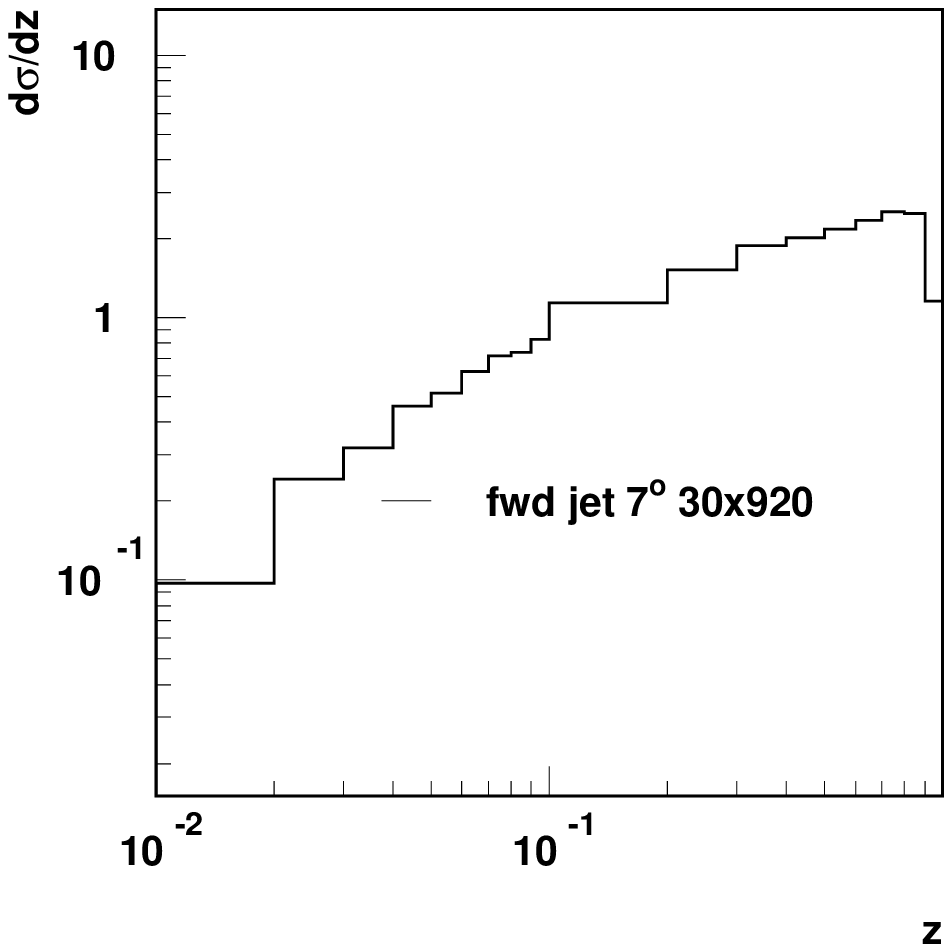,height=8.cm}
    \end{minipage}
  \end{center}
\caption{\it The values of the splitting variable $z$ for events
  satisfying the shown forward jet criteria, with $\theta = 7^o$ at
  HERA energies. }
    \label{zval_fjet_hera} 
\end{figure}

The basic event selection criteria 
for forward jet production at HERA are given in
Fig.~\ref{zval_fjet_hera}.
\par
The criterion $E^2_\Prp/Q^2$ is essential to suppress
the DGLAP contribution within a typical range in $x$  
of $10^{-3} < x < 10^{-2}$.
The evolution takes place from the large $x_{jet}$ down to the small
$x$ with a typical range at HERA energies of $\Delta x=x/x_{jet} >
0.01$.  In order to justify the use of an evolution equation
(instead of a fixed order calculation) one would require at least 2 or
more gluon emissions during the evolution. To roughly estimate the
energy fractions $z_i$ of 3 gluon emissions between $ 10^{-3} < x <
10^{-1}$, one can assume that all gluons carry equal energies. Then
the range of $\Delta x \sim 0.01$ results in $z \sim 0.2 $, which is
far from being in the very small $z$ region, where the BFKL or CCFM
approximations (treating only the $1/z$ terms in the gluon splitting
function) are expected to be appropriate. In Fig.~\ref{zval_fjet_hera}
we show the values of the splitting variable $z$ in events satisfying
the forward jet criteria at HERA energies obtained from the Monte
Carlo generator \CASCADE~\cite{jung_salam_2000}.  Since the values of
the splitting variable $z$ are indeed extending into the medium and
large $z$ region (the majority has $z>0.1$), it is questionable,
whether the BFKL evolution equations, including only the $1/z$
part of the gluon splitting function, are applicable.
CCFM also has a $1/(1-z)$ term, which  
means that in any case one expects quite a few emissions for $z \to  1$, but
still, the medium $z$ range is questionable.
\par 
The implementation of the full
DGLAP splitting function into CCFM is problematic.
Naively one would simply replace
\begin{equation}
  \frac1{1-z} \to \frac{1}{1-z} - 2 + z(1-z)\,,
\end{equation}
in the CCFM splitting function. However doing this leads to negative
branching probabilities, so one needs to be a little more `subtle'.
\par
For example a positive definite branching probability can be obtained
by making the following replacements
\begin{equation}
  \frac{1}{z} \to \frac{1-z}{z} + Bz(1-z)\,, \qquad 
  \frac{1}{1-z} \to \frac{z}{1-z} + (1-B)z(1-z)\,,
\end{equation}
resulting in: 
\begin{eqnarray}
P(z,q,k)& = &\asb \left(\kt^2\right) \left( \frac{(1-z)}{z} + (1-B)z(1-z)\right)
\Delta_{ns}(z,q,k) \\
& &  + \asb\left((1-z)^2q^2\right) \left(\frac{z}{1-z} + Bz(1-z)\right)
\nonumber
\end{eqnarray}
where $B$ is a parameter to be chosen arbitrarily between $0$ and $1$. 
As a consequence of the replacement, the Sudakov form factor will change to:
\begin{equation}
\log\Delta_s(\Pmax^2,Q_0^2) =
 - \int_{Q_0^2} ^{\Pmax^2}
 \frac{d q^{2}}{q^{2}} 
 \int_0^{1-Q_0/q} dz \alphasb\left(q^2(1-z)^2\right)
 \left( \frac{1-z}{z'} + (1-B)z(1-z) \right)
\end{equation}
and also the non-Sudakov form factor needs to be replaced by:
\begin{equation}
\log\Delta_{ns} =  -{\bar \alpha}_s\left(\kt^2\right)
                  \int_0^1 dz'
                        \left( \frac{1-z}{z'} + (1-B)z(1-z) \right)
                        \int \frac{d {q'}^2}{{q'}^2} 
              \Theta(k-q')\Theta(q'-z'q)
\end{equation}
which can be rewritten
\begin{equation}
\log\Delta_{ns} =  -{\bar \alpha}_s\left(\kt^2\right)
                  \int_0^1 dz'
                        \left( \frac{1-z}{z'} + (1-B)z(1-z) \right)
                        2\ln \frac{k}{z'q} \Theta(k - z' q)
\end{equation}
 
\par
The proposed changes
to the original CCFM splitting function are `hacks'. There are reasonable
arguments, to change the scale in $\alpha_s$ to $q^2(1-z)^2$ everywhere,
to include the full splitting function instead of only the singular parts and
also to impose the kinematic constraint, with its consequences. These `hacks'
are of course constrained by the requirement that the NLL corrections
corresponding to the given branching equation are similar to the true
NLL corrections.
However, this will not be the real final solution and 
perhaps it is a need for a better starting point,
embodying both angular ordering and symmetry right from the start.

\section{\boldmath Generators for \sx\ evolution}
\label{sec:gener-sc-evol}
Three different Monte Carlo event generators are available,
specifically devoted to small $x$ processes (in chronological order):
\SMALLX\ ~\cite{\SMALLXMC}, \ldcmc\ ~\cite{\LDCMC} and \CASCADE\ 
~\cite{\CASCADEMC}

\subsection{\SMALLX}
\label{sec:smallx}
\SMALLX ~\cite{\SMALLXMC} is a Monte Carlo event program which generates
events on parton level at small $x$ in $ep$ scattering. It uses 
the CCFM~\cite{\CCFM} evolution equation for the initial state
cascade convoluted with the \kt-factorized off-shell matrix elements
of ~\cite{CCH} for light and heavy quark pair production.
Modifications of the original \SMALLX ~\cite{\SMALLXMC} version
concerning the non-Sudakov form factor $\Delta_{ns}$ were necessary to
obtain a reasonable description of the structure function $F_2(x,Q^2)$
as well as hadronic final states like the forward jets at
HERA~\cite{smallx_f2,CASCADE}.
\par
In \SMALLX\ the initial state gluon cascade is generated in a forward
evolution approach. The gluon evolution starts from the proton side
with an initial gluon distribution according to (including a Gaussian
intrinsic \kt\ distribution around $k_0$):
\begin{equation}
  x_0 G_0(x_0,\kti{0}^2) =
  N \cdot (1-x_0)^4 \cdot \exp{\left(-\kti{0}^2/k_0^2\right)}
\end{equation}
with $N$ being a normalization constant. A gluon with momentum
fraction $x_i$ and transverse momentum \kti{i} is allowed to branch
into a virtual ($t$-channel) gluon with momentum $k_{i+1}$ 
and a final state gluon with momentum
$p_{i+1}$ according to the CCFM splitting function \cite{\CCFM}.
This  procedure is repeated until the
next emitted gluon would violate the angular bound $\Pmax$ given by
the matrix element.
\par
\SMALLX\ generates the full parton level structure, but due to the
complicated structure of the initial state branchings and the phase
space of the matrix element, a weight is associated with each event.
Although \SMALLX\ produces weighted events and  therefore is
inefficient for generating specific exclusive signatures, it can be
used for the CCFM evolution and to calculate the inclusive structure
function $F_2(x,Q^2)$. By performing a fit to measurements of the
inclusive structure function $F_2(x,Q^2)$, \SMALLX\
 can be used to determine
the unintegrated CCFM gluon density $x {\cal A}(x,\kt^2,\Pmax^2)$ in a
grid in $x$, $\kt ^2$ and $\Pmax^2$ (with $\Pmax$ being the scaled maximum
angle allowed for any emission with $\Pmax^2 = x^2_{n-1} \Xi s$).  This
numerical representation can be used for any other calculation. It is
advantageous, that the full parton level is generated during the
evolution, since this information can be used for comparison with
other event generators adopting a more efficient backward evolution
approach~\cite{\CASCADEMC}.
\subsection{\CASCADE}
\label{sec:cascade}
\CASCADE ~\cite{\CASCADEMC} is a full hadron level event generator
which uses \kt-factorization of the cross section into an
off-shell matrix element and an unintegrated gluon density function.
The initial state cascade is generated in a backward evolution approach, which
was necessary for an efficient event generation. 
The Lund string fragmentation package \jetset /\pythia ~\cite{\PYTHIAMC} is
used for hadronization.  \CASCADE\ can be used in $ep$, $\gamma p$ and
also $p\bar{p}$ processes.
\par
The hard scattering process is calculated using the off-shell matrix
elements given in~\cite{CCH} for light and heavy quark pair
production, or $\gamma g \to J/\psi g$~\cite{saleev_zotov_a}
convoluted with the unintegrated gluon density 
$x {\cal A}(x,\kt^2,\Pmax^2)$.  
It could be shown in~\cite{\CASCADEMC} that a backward
evolution approach~\cite{\CASCADEMC,PYTHIAPSa,LEPTOPS} is possible for
small $x$ processes, which are not restricted to strong \kt\ ordering
in the initial state cascade. This was the main ingredient for the
development of a time efficient Monte Carlo event generator.
\par
The backward evolution starts from the hard scattering process and
evolves the partons {\it backwards} towards the proton. This approach
is much more efficient, compared to the strategy of
\SMALLX\, which uses a forward
evolution approach. However, it requires the unintegrated gluon
density to be determined beforehand. The CCFM unintegrated gluon
distribution has been determined from a Monte Carlo solution of the CCFM
evolution equation which has been fitted to the
measured structure function $F_2(x,Q^2)$.  It is advantageous to have
the unintegrated gluon density determined in a Monte Carlo approach
with full control over the partonic state, as available in \SMALLX,
since this allowed to show~\cite{jung_salam_2000} that the backward
evolution produces identical results to the forward evolution approach
on the parton level.  The proof of equivalence of the forward and
backward evolution is a unique feature of CCFM. 
For DGLAP this has only been shown at at an inclusive level, while at the
exclusive level, complications can arise related to differences in the
treatment of angular ordering between forward and backward evolution.
\par 
The program code is available from
\verb+http://www.quark.lu.se/~hannes/cascade+
\subsection{\ldcmc}
\label{sec:ldcmc}
Since the LDC model is inherently forward--backward symmetric, it is
natural to design a Monte Carlo in the same way. In the \ldcmc\ 
program all emissions are generated in one go from an incoming,
non-perturbative gluon with energy fraction $x_0$, using a generating
function
\begin{eqnarray}
  \label{eq:ldcgen}
  G(a=\log{\frac{Q^2 x_0}{\kti{0}^2 x}},
  b=\log{\frac{x_0}{x}}) &=& \sqrt{\frac{\alb a}{b}}I_1(2\alb\sqrt{ab}) =
  \sum_{n=1}^\infty\frac{\alb^na^nb^{n-1}}{n!(n-1)!}\\
  &=&\sum_{n=1}^\infty\int\alb^n\Pi_j
  \frac{dz_{j+}}{z_{j+}}\frac{dz_{j-}}{z_{j-}}\delta(\log{\frac{x}{x_0}}-\sum_j\log{z_{j+}}).
  \nonumber
\end{eqnarray}
After this, the azimuthal angles of each emission are selected according
to a flat distribution and a number of correction factors are included
to produce a weight for the generated partonic state. \ldcmc\ can
either produce weighted events or use the weight as a hit-or-miss
probability to produce unweighted events. The weight can include many
things such as:
\begin{itemize}
\item The full splitting functions (also for quark propagators)
  instead of the simplified
  $\frac{dz_{j+}}{z_{j+}}\frac{dz_{j-}}{z_{j-}}$. Optionally the full
  matrix element for a $2\rightarrow2$ sub-collision can be used. In
  particular for the quark box closest to the virtual photon the full
  off-shell matrix element can be used. 
\item The standard Sudakov form factors.
\item Emissions not satisfying the LDC constraint in
  eq.(\ref{eq:ldccut}) are given zero weight.
\item The running of \as. The scale is taken to be the transverse
  momentum of the emitted parton which, due to the LDC constraint, is
  always close to the highest scale in the emission.
\end{itemize}
\par
Averaging over weights for a given $x_0$, $x$ and $Q^2$, it is then
possible to fit the input gluon (and quark) distribution(s) to describe
eg.\ $F_2(x,Q^2)$.
\par
Using these input distributions, parton-level events can be generated.
Final-state parton cascades are added using the colour-dipole cascade
implemented in \ariadne\ for $e^+e^-$ annihilation but only allowing
emissions which are below the LDC constraint in eq.(\ref{eq:ldccut}).
Finally, standard Lund string fragmentation can be added to produce
final-state hadrons.
\par
\ldcmc\ is distributed together with the \ariadne\ program, available
from\\
\verb|http://www.thep.lu.se/~leif/ariadne| 
but in using it, one should keep in mind, that the reproduction of
data on eg.\ forward jets is very poor. The problem can be traced to
the non-singular parts of the gluon splitting function. In a newer
(not yet released) version, it is possible to allow only gluonic
chains and only the singular parts of the gluon splitting function.
The results are then consistent with what is obtained with \SMALLX\ 
and \CASCADE.
\begin{table}
\begin{tabular}[htb]{|l|c|c|c|c|}
\hline
Name & QCD cascade & applicable & processes & event record \\
\hline
\SMALLX  & forward evolution & $ep$ & 
$\gamma^* g^*\to q\bar{q}$ & parton level\\
\protect\cite{\SMALLXMC} & with CCFM & & $\gamma^* g^* \to Q\bar{Q}$ & weighted events \\
 \hline
\CASCADE & backward evolution & $ep$,$\gamma p$,
$p\bar{p}$& $\gamma^* g^* \to q\bar{q}$ & parton level\\
\protect\cite{\CASCADEMC}
& with CCFM & & $\gamma^* g^*  \to Q\bar{Q}$ & unweighted events \\
& using unintegrated & & $\gamma g^*  \to J/\psi g $ & hadronization  \\
& gluon density     & & $g^* g^*  \to q\bar{q}$ & 
via \jetset /\pythia ~\protect\cite{\PYTHIAMC} \\
&      & & $g^* g^*  \to Q\bar{Q}$ & \\
\hline
\ldcmc & forward--backward & $ep$ &
$\gamma^* g^*\to q\bar{q}$ & weighted or \\
\protect\cite{\LDCMC}
& symmetric LDC &  & $\gamma^* g^*  \to Q\bar{Q}$ & unweighted events  \\
& evolution & & & final state cascade \\
& & & & hadronization \\
& & & & via \jetset /\pythia ~\protect\cite{\PYTHIAMC}\\
\hline
\end{tabular} 
\caption{{\it Summary and overview over existing Monte Carlo event generators for
small $x$ physics.}}
\end{table}

\section{\boldmath Conclusions}

In this summary report we presented the state of the art of small $x$ physics in
the year 2001. Significant progress has been made in the understanding of
the small $x$ evolution equations, CCFM and BFKL. It has been possible for the
first time to describe the structure fucntion $F_2(x,Q^2)$ and also hadronic
final state measurements, like forward jet production, with the CCFM evolution
equation implemented into a Monte Carlo program. In more detailed studies the
need for improving the small $x$ splitting functions became evident,
also from considering next-to-leading corrections to the BFKL equation.
However, considering these improvements in detail, 
it became also clear, that this is not a trival task.
\par
Significant progress has also been made in the understanding of
$\kt$-factorization in general and the calculation of the off-shell matrix
elements. In certain approximations the off-shell matrix elements are already
calculated to order ${\cal O}(\alpha_s^2)$. However, some critical points still
need to be clarified, as the gauge invariance of the $\kt$-factorization
approach is not yet clear, if considered beyond leading order.
\par
For the first time, a comparison of all available paramerisations of
unintegrated gluon distributions was made, showing significant differences, which
indicate, that more 
(and  more exclusive) measurements need to be used in constraining the
unintegrated gluon distributions further. For the first time, we are in a position
to try to perform a global fit, similar to those using the collinear
approximation, to determine the unintegrated gluon density.
\par
Our understanding of \sx\ physics is far from complete. There are a
number of theoretical and phenomenological issues which need  to be further
settled. In this overview we have mentioned many such issues. On the
theoretical side, we need to understand the convergence of the
perturbative expansion of the evolution kernel, and we also need to
calculate the impact factors to next-to-leading order. On the
phenomenological side it is important to understand the importance of
non-singular terms in the gluon splitting function especially when
implemented in event generators. It is also important to understand
the uncertainties involved with fitting unintegrated parton distributions
to data. These parameterizations should then be compared to a wide
variety of data, possibly requiring the calculation of new off-shell
matrix elements for other than the mentioned processes.
\par
Finally, we want to note that not all aspects of \sx\ physics have
been covered in this paper. Phenomena such as rapidity gaps, shadowing
effects and multiple interactions are also interesting aspects of \sx\ 
evolution, and will be included in forthcoming meetings and publications.

\section*{\boldmath Appendix: The \sx\ collaboration}
\label{sec:sx-collaboration}
Today the work on these issues is spread out on a number of different
small groups around the world. Although some of these group are
already collaborating on an informal basis, 
it was agreed in the meeting, that 
the future work could be more coordinated. As a consequence of this
 a \emph{Small-$x$ Collaboration} was formed.
The idea is to start small and informal and to set up a web site an a
mailing list, but also to organize small meetings such as the one held
in Lund resulting in this paper.
\par
The web site is located at \verb|http://www.thep.lu.se/Smallx| and
will among other things contain a compilation of subroutines
implementing parameterizations of unintegrated parton distributions and
off-shell matrix elements and compilations of relevant papers,
theoretical as well as phenomenological and experimental. The mailing
list would be used to announce new results, to ask for help from
experts in the field and so on.

The Small-$x$ Collaboration will, of course, be open to anyone in the
field. To join the collaboration one can simply join the mailing list
by sending a mail to \verb|smallx-subscribe@thep.lu.se| (more
instructions can be found on the web site).

\raggedright

\end{document}